\documentclass[a4paper, fleqn, review]{cas-sc}

\usepackage{cleveref}
\usepackage{textcomp}
\usepackage{makecell}
\usepackage[numbers, sort&compress]{natbib}
\usepackage{bm} 

\def\tsc#1{\csdef{#1}{\textsc{\lowercase{#1}}\xspace}}
\tsc{WGM}
\tsc{QE}

\newdefinition{rmk}{Remark}
\newproof{pf}{Proof}
\newproof{pot}{Proof of Theorem \ref{thm}}

\begin{document}

\let\WriteBookmarks\relax
\def\floatpagepagefraction{1}
\def\textpagefraction{.001}

\shorttitle{ADfiRe: An empirical- and CFD-informed advection-diffusion-reaction thermal model for wildfire spread}

\shortauthors{Vogiatzoglou, et al.,}   

\title [mode = title]{ADfiRe: An empirical- and CFD-informed advection-diffusion-reaction thermal model for wildfire spread}  

\author[1]{Konstantinos Vogiatzoglou}[orcid=0009-0004-8333-247X]\cormark[1]
\cortext[cor1]{Corresponding author}

\ead{kvogiatzoglou@uth.gr}

\credit{Conceptualization, Data curation, Formal analysis, Investigation, Methodology, Resources, Software, Validation, Visualization, Writing – original draft, Writing – review and editing}

\author[1]{Nikolaos Sdiros}

\credit{Conceptualization, Investigation, Software, Writing – review and editing}

\author[1]{Costas Papadimitriou}

\credit{Conceptualization, Project administration, Supervision, Writing – review and editing}

\author[1]{Konstantinos Ampountolas}

\credit{Conceptualization, Writing – review and editing}

\author[2]{Petros Koumoutsakos}

\credit{Conceptualization, Writing – review and editing}

\author[1]{Vasilis Bontozoglou}

\credit{Conceptualization, Formal analysis, Investigation, Methodology, Project administration, Supervision, Validation, Writing – original draft, Writing – review and editing}

\affiliation[1]
            {organization={Department of Mechanical Engineering, University of Thessaly},
            address={Leoforos Athinon, Pedion Areos,},
            city={Volos 38334},
            country={Greece}}

\affiliation[2]
            {organization={Computational Science and Engineering Laboratory, Harvard University},
            address={19 Oxford St,},
            city={Cambridge, MA 02138},
            country={United States}}

\begin{abstract}
Real-time wildfire forecasting requires a trade-off between rapid yet physically relevant models and high fidelity but computationally costly models of flame spread over heterogeneous, multiphase solid fuel. The present study introduces ADfiRe, a hybrid framework that integrates empirical correlations and CFD-informed parameterizations into a physics-based advection-diffusion-reaction (ADR) formulation. The key  features of ADfiRe are the use of established fuel databases, the capacity to handle multiple fuels, and a methodology to capture the effect of arbitrary topography. Furthermore, it includes the effect of flame tilting due to  wind, thus accounting for the differences between plume- and wind-dominated fire propagation. ADfiRe resolves explicitly the coupled thermochemical degradation as well as heat and mass transport mechanisms of the advancing fire front. The wind  field is modeled based on an appropriately defined Byram number. The model is validated against five well-documented experimental field-scale fires in Mediterranean shrublands, showing quantitative agreement with key metrics such as the rate of spread (ROS) and the fireline intensity, through a two-stage, Arrhenius-based heterogeneous reaction scheme coupling solid-fuel pyrolysis and oxygen-limited char/volatile combustion.
\end{abstract}

\begin{graphicalabstract}
    \includegraphics[width=1.0\textwidth]{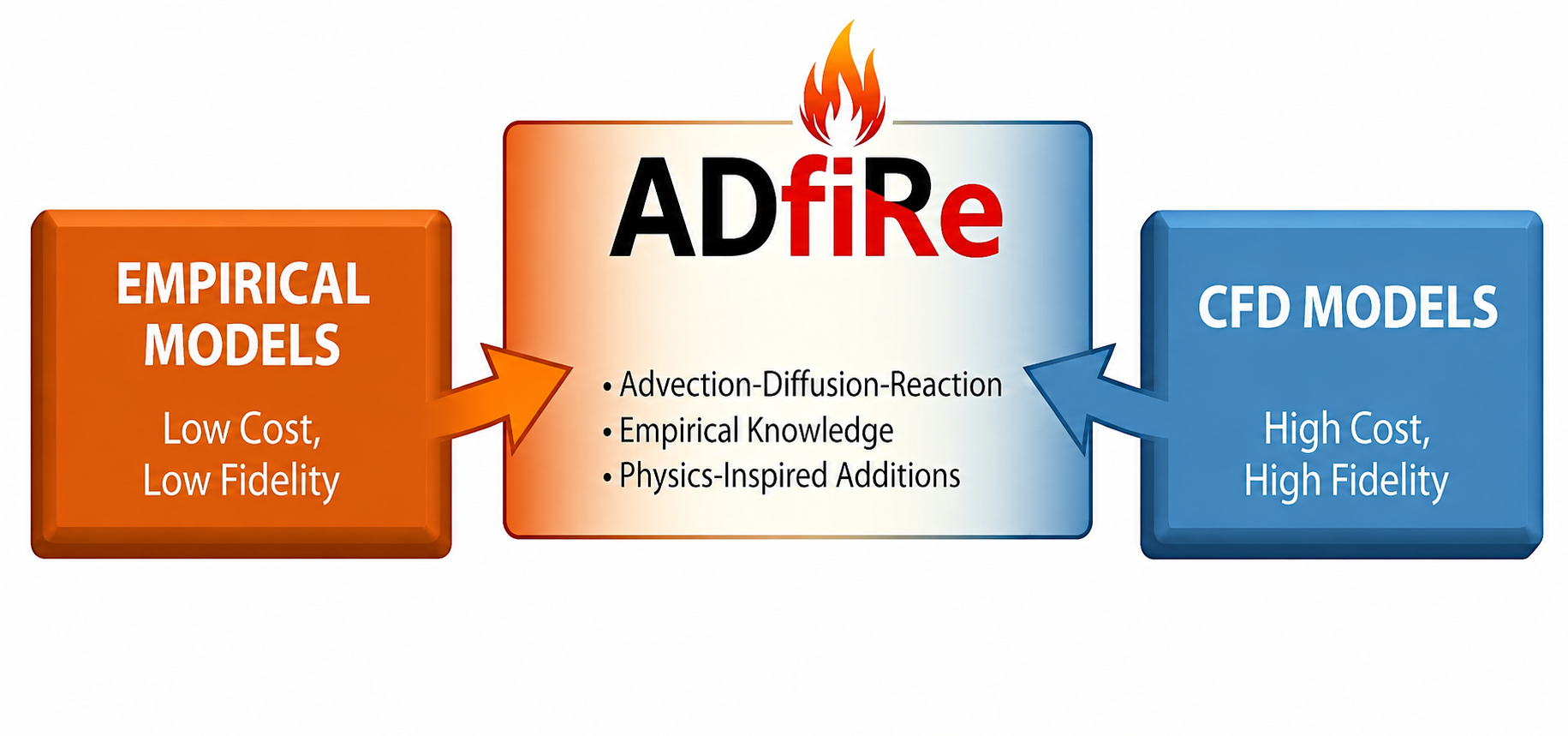}
\end{graphicalabstract}

\begin{highlights}
    \item {Merges empirical formulations with CFD-resolved quantities via ADR framework.}

    \item {Accounts for multiple fuel components with distinct thermophysical properties}.
    
    \item {Couples terrain slope and wind-driven flame deflection via modified Byram number.} 
    
    \item {Predicts ROS and fireline intensity within 5-13\% error across five wildland fires.}

    \item {Achieves sub-two-minute runtimes on a single CPU, enabling operational use.}
\end{highlights}


\maketitle

\section{Introduction}
    \label{sec:Introduction}

The global intensification and accelerating frequency of wildfire events worldwide have resulted in prolonged fire seasons and unprecedented annual burned areas \cite{Abatzoglou_2016, Giannaros_2022, Jain_2022, Giannaros_2023,Cunningham_2024}.The escalating wildfire threat underscores the critical need for accurate and reliable fire propagation models to enable timely evacuation and effective suppression strategies, protect communities at the wildland–urban interface (WUI), and reduce potentially devastating socioeconomic losses \cite{Thomas_2017,Hessburg_2021}.

Wildland flame spread involves multiscale and multiphysics phenomena driven by the interaction between thermochemical degradation \cite{Seron_2005, Grasso_2020, Vogiatzoglou_2024, Montilla_2025, Simeoni_2011, Rodman_2007} and turbulent transport mechanisms \cite{Finney_2015, Viegas_2012, Viegas_2006}. Front advancement is primarily governed by transient convective, diffusive, and radiative heat fluxes that are continuously modulated by flame dynamics \cite{Finney_2015}, fuel variability \cite{Banerjee_2020}, topography \cite{Eftekharian_2019}, and atmospheric conditions \cite{Sun_2009}. Significant non-linear fluid dynamic effects (such as buoyancy-driven updrafts and shear-induced instabilities \cite{Sanjuan_2014}) alter fire behavior, challenging the adequacy of traditional steady-state models and suggesting the development of more advanced frameworks that integrate time-dependent physics and atmospheric boundary layer analysis alongside the traditional fire-environment triangle (fuel, weather, topography).

Typical methodologies towards wildfire modeling range between two computational extremes \cite{Pastor_2003, Rothermel_1972, Pimont_2012}. At one end, empirical models provide rapid macroscopic estimations of the rate of spread (ROS) and fireline intensity by mapping averaged environmental conditions to experimental data. Such models remain the backbone of modern operational prediction systems due to their low computational cost, which permits real-time deployment. Rothermel’s formulation \cite{Rothermel_1972}, iteratively refined through extensive observational data \cite{Andrews_2018}, drives primary decision-support tools such as BehavePlus \cite{Andrews_2014}, FlamMap \cite{Finney_2006}, and FARSITE \cite{Finney_1998}. A critical legacy of these empirical frameworks is the establishment of standardized fuel behavior databases \cite{Anderson_1982, Scott_2005}, which compile the structural and thermochemical parameters necessary to characterize the combustion dynamics of diverse vegetation types.

At the other end, modern computational fluid dynamics (CFD) simulators, such as FDS \cite{Mell_2007, Mell_2009}, FIRESTAR \cite{Morvan_2006}, FIRETEC \cite{Linn_2002, Pimont_2009, Canfield_2014}, offer detailed insights into fire evolution mechanisms, albeit through lengthy and costly computations that preclude real-time operational execution.  For example, they reveal differences in flow-field characteristics between plume- and wind-driven fire regimes, as well as resolve the emergence of secondary flow structures that contribute to fire propagation \cite{Finney_2015,Finney_2022, Morvan_2018, Linn_2005, Shamsaei_2023, Frangieh_2020}. Still, no single simulator can resolve all relevant spatiotemporal scales, and researchers target specific resolutions (i.e. microscopic domains for combustion kinetics, intermediate meshes for turbulent momentum transfer, or landscape-scale grids for macroscopic fire-atmosphere feedback \cite{Morvan_2011}).

Intermediate-complexity models, known as advection-diffusion-reaction (ADR) frameworks \cite{Vogiatzoglou_2024, Montilla_2025, Burger_2020}, are promising candidates for bridging the gap between the above two extremes. By respecting the fundamental mechanisms of coupled heat and mass transfer, alongside simplified reaction kinetics, they offer physical fidelity, while their modest computational cost permits real-time forecasting. Yet, their operational efficiency relies on the precise calibration of a restricted parameter set and, as a result, they are susceptible to both epistemic (lack of precise modeling knowledge \cite{Kelly_2025}) and aleatoric (inherent environmental variability \cite{Hilton_2015}) uncertainty. Validation with field data from well-documented fires is an imperative first step for building confidence in such models.

The present study introduces ADfiRe, a physics-based ADR framework that significantly expands upon our prior research \cite{Vogiatzoglou_2024}. In particular, ADfiRe (i) uses established fuel databases \cite{Anderson_1982, Scott_2005} with the additional capacity to handle multiple fuels with different burning characteristics, (ii) extends Nelson’s methodology \cite{Nelson_2002, Nelson_2015} to the ADR framework, capturing the effect of arbitrary two-dimensional (2D) topography, and (iii) predicts the effect of flame tilting due to the wind and thus accounts for the differences between plume- and wind-dominated fire propagation \cite{Morvan_2018, Morandini_2010, Trouve_2020, Morvan_2025}. The latter is achieved by heuristically modifying the flow field based on an appropriately defined Byram number ($N_c$), in accordance with the findings of recent CFD studies \cite{Pimont_2012, Shamsaei_2023, Frangieh_2020}. Last but not least, the predictions of ADfiRe are validated by successful comparison with five sets of well-documented field-scale data \cite{Nelson_2015, Morandini_2010}.

Compared to recent black-box modeling efforts based on deep learning architectures, such as convolutional and recurrent neural networks (ConvLSTMs) \cite{Marjani_2024, Wang_2024, Zhou_2025} as well as transformers, physically interpretable models like ADfiRe can provide a reliable foundation for parameter calibration, data assimilation, and uncertainty-aware simulation \cite{Vogiatzoglou_2025} due to their grounding in fundamental conservation laws. In turn, these capabilities lay the essential groundwork for more accurate fire risk quantification and effective safety management.

The remainder of this manuscript is structured as follows. Section~\ref{sec:Model_Derivation} details the analytical derivation and physical parameterization of the ADfiRe framework. Section~\ref{sec:Results} presents a comprehensive validation of the proposed simulator against five experimental field-scale fires. Finally, Section~\ref{sec:Conclusions} summarizes the key findings and outlines potential directions for future research. 

\section{Problem setup and model derivation}
    \label{sec:Model_Derivation}

We first outline the model's primary components, with the remainder of the section describing each in detail. 

\subsection{Model outline}
    \label{subsec:Model_Outline}

\begin{figure}
    \centering
    \includegraphics[width=.45\textwidth]{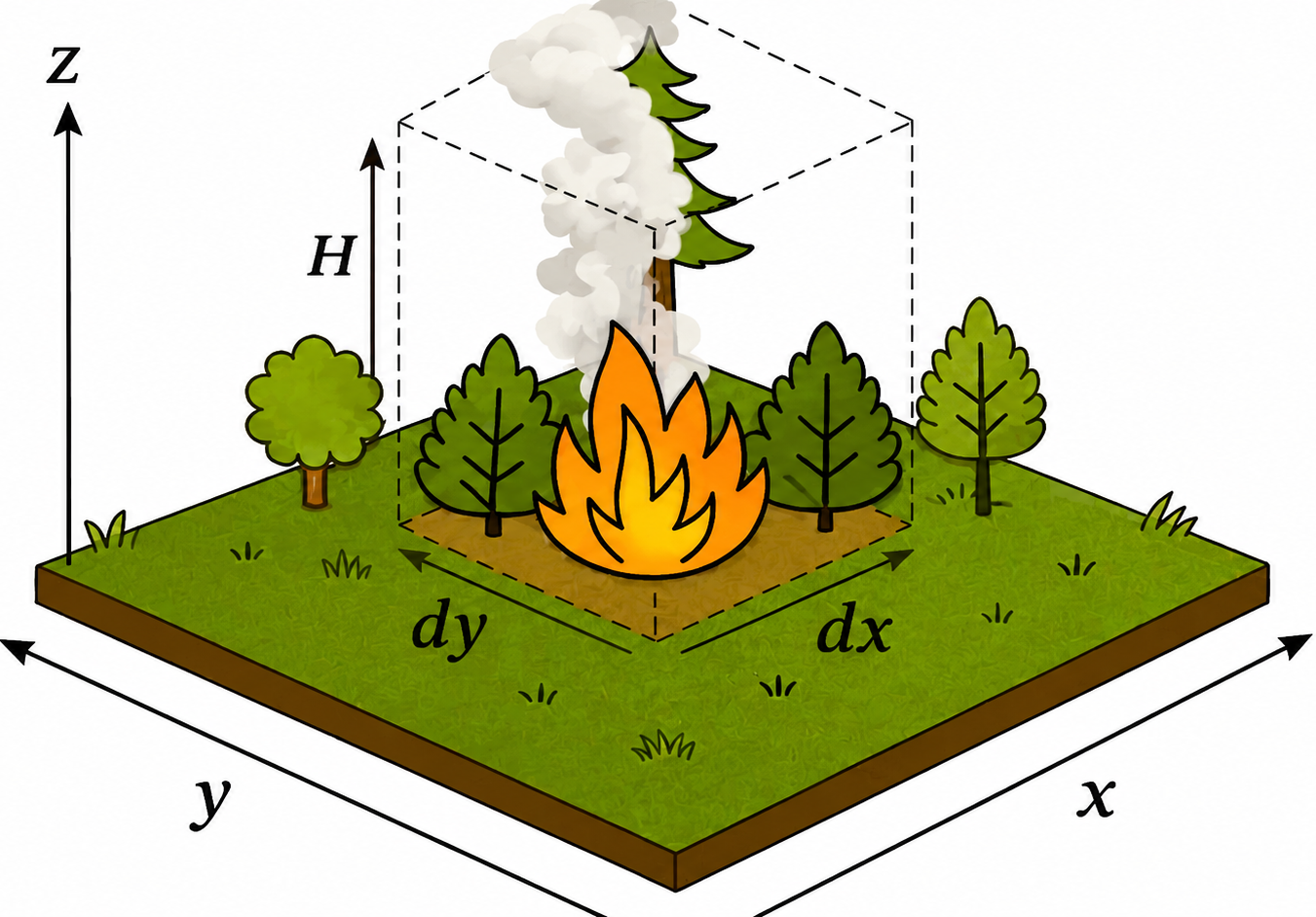}
    \caption{Schematic of the differential control volume $(dx, dy, H)$ used to formulate the vertically integrated energy balance across the terrain and vegetation layer.}
    \label{fig:Figure_1}
\end{figure}

The spatial domain is defined by a Cartesian coordinate system $(x, y, z)$, where the complex landscape is mapped by the terrain elevation, $z = H_t(x, y)$ [\text{m}], and by the local vegetation height above ground, $H_v(x, y)$ [\text{m}]. Ambient wind conditions are characterized by a reference wind speed, $\mathbf{u}_{\rm{w}}$ [m s$^{-1}$], measured at the standard meteorological height of 10 \text{m} above the surface. To formulate the energy balance, a differential control volume $(dx, dy, H)$ is defined, spanning a horizontal area ($A = dx \, \times \, dy$) and encompassing the entire vertical depth $H$ of the maximum fuel bed (see Fig.~\ref{fig:Figure_1}). Variations in canopy bulk density along the $z$-direction are neglected; thus, every distinct fuel species $i$ is described by its initial surface load $m_{0, i}$ [$\text{kg} \, \text{m}^{-2}$], defined as the total solid mass per unit of surface area. The assumption of a uniform bulk density appears reasonable for canopies like the Mediterranean shrublands presently considered \cite{Kutchartt_2024, Morvan_2001}. However, it would be questionable for some situations, for example a forest of tall and densely foliated trees \cite{Marcozzi_2025}.

Within each control volume, the thermodynamic state is defined by the temperature field $T(x, y, t)$ [\text{K}]. This formulation relies on two primary physical simplifications, standard in reduced-order wildfire modeling. First, the temperature is taken as uniform in the vertical ($z$) direction, a valid approximation given that the depth of the fuel bed is typically much smaller than the macroscopic horizontal scales of the advancing fire front \cite{Montilla_2025, Grasso_2020, Burger_2020}. Second, local thermal equilibrium is assumed between the solid fuel and the gaseous phase \cite{Asensio_2023}.

The assumption of thermal equilibrium is considered reasonable at the current level of approximation  \cite{Asensio_2023, Mandel_2007}, given that the model reaction kinetics also does not differentiate between the solid and the gaseous phase. Based on the satisfactory prediction of macroscopic measurables such as the rate of spread and the flame intensity, it is argued that the "averaged" description proposed is a useful simplification in the present context.

The energy balance incorporates two source terms: an endothermic sink representing fuel moisture evaporation and volatiles release (pyrolysis) and an exothermic term representing volatiles and solid fuel combustion (oxidation). The exothermic reaction kinetics include explicit dependence on the concentration, $C_{bulk}$ [$\text{kg} \, \text{m}^{-3}$], of oxygen in the bulk of the flame, with $C_{bulk}$ estimated by comparing the entering fresh air with the flue gasses produced by combustion.

Heat transport across the control volume is governed by  advective, diffusive, and radiative mechanisms \cite{Nelson_2015}. The advective contribution is modeled by the effective mean velocity of the intra-canopy gas phase, $\mathbf{u}_{\textnormal{eff}}(x, y)$ [$\text{m} \, \text{s}^{-1}$]. The rest are modeled as short-range interactions (i.e., turbulent-buoyant dispersion, wind-induced mixing, and dense-medium radiation) and are lumped into a spatially varying effective diffusion coefficient, $\mathbf{D}_{\textnormal{eff}}(x, y)$ [$\text{m}^2 \, \text{s}^{-1}$]. Finally, vertical thermal losses to the ambient are quantified by an overall loss coefficient, $U$ [$\text{W} \, \text{m}^{-2} \, \text{K}^{-1}$].

In line with our previous work \cite{Vogiatzoglou_2024} and with the 2D nature of the model, no attempt is made to directly solve the fluid mechanics problem, and the flow field needed in the energy balance is provided by heuristic arguments. To account for flame inclination relative to the ground (and extending an idea originally proposed by Nelson \cite{Nelson_2002}), the mean effective velocity $\mathbf{u}_{\mathrm{eff}}$ that governs advective heat transport is taken to comprise two components: the velocity $\mathbf{u}_{\mathrm{shear}}$ induced by ambient wind shear and $\mathbf{u}_{\mathrm{proj}}$, an appropriate projection along the ground of the buoyant velocity vector. Thus,  $\mathbf{u}_{\mathrm{eff}} = \mathbf{u}_{\mathrm{shear}} + \mathbf{u}_{\mathrm{proj}}$. The buoyant velocity vector is inclined relative to the ground, not only because of the local terrain slope $\varphi$ [$^{\circ}$], but also because the flame may be tilted by the wind. The tilt angle, $\Phi$ [$^{\circ}$], is correlated with the local Byram number $N_c$, based on the data of Moradini and Silvani \citep{Morandini_2010}. For one-dimensional (1D) propagation in the direction of the wind, the magnitude of $\mathbf{u}_{\mathrm{proj}}$ depends simply on the sum  $(\varphi+\Phi)$. However, for 2D topography and arbitrary wind direction, accurate calculation of $\mathbf{u}_{\mathrm{proj}}$ necessitates vector transformations, as detailed in Appendix A2 (Supplementary Material).

\subsection{Fuel mixture and thermochemical reaction kinetics}
    \label{subsec:Fuel_Kinetics}

Wildland fuel complexes exhibit profound geometric and thermochemical heterogeneity \cite{Andrews_2018}. The resulting solid-gas combustion problem is intrinsically heterogeneous and multiphase, with reaction rates governed jointly by Arrhenius kinetics and interfacial oxygen mass transfer. To capture this diversity, ADfiRe parameterizes vegetation dynamics utilizing established fuel catalogs \cite{Anderson_1982, Scott_2005}. The fuel bed is modeled as a mixture of $N$ discrete components ($i=1, \dots, N$), all with the same solid bulk density, $\rho$ [$\text{kg} \, \text{m}^{-3}$], but each characterized by its specific surface-to-volume ratio (SAV), $\sigma_i$ [$\text{m}^{-1}$].

The initial surface loading of each fuel component, $m_{0, i}$, is defined as the sum of its constituent moisture, $m_{0 \alpha, i}$, and dry combustible material, $m_{0 \beta, i}$, such that $m_{0, i} = m_{0 \alpha, i} + m_{0 \beta, i}$. As thermal degradation progresses, the transient residual masses of these fractions at any time $t$ are denoted as $m_{\alpha, i}$ and $m_{\beta, i}$, yielding a time-dependent total mass $m_i = m_{\alpha, i} + m_{\beta, i}$. For notational purposes, quantities summed across all $N$ fuel components are denoted without the subscript $i$ (e.g., the total initial surface load $m_0 = \sum_{i=1}^{N} m_{0, i}$, and total transient mass $m = \sum_{i=1}^{N} m_i$).

The properties of the fuel bed are then derived from these constituent masses. The initial leaf area index, $\text{LAI}_0$, is formulated as $\text{LAI}_0 = (1 / \rho) \sum_{i=1}^{N} m_{0, i} \sigma_i$. The total packing ratio, $a$, representing the volume fraction occupied by the solid phase, is given by $a = m_0 / (\rho H)$. Finally, the mass of the gaseous phase per unit ground area, $m_g$ [$\text{kg} \, \text{m}^{-2}$], is assumed to remain approximately constant and is evaluated as $m_g \approx \rho_g (H - m_0/\rho)$, where $\rho_g$ [$\text{kg} \, \text{m}^{-3}$] represents a reference gas density at the mean flame temperature.

The thermal degradation of the solid fuel is formulated as a two-stage process \cite{Baranovskiy_2022, Liodakis_2005, Park_2010}, with the respective reaction rate of each stage defined below per unit of horizontal surface area rather than on a volumetric basis. The first stage, encompassing the evaporation of fuel moisture and the release of volatiles, is strictly endothermic, and its thermal effect is approximated by the latent heat of evaporation \cite{Leckner_1999}. The respective reaction rate is described by first-order Arrhenius kinetics, expressed for each fuel component $i$ as:
\begin{equation}
    \label{eq:Endothermic_Reaction}
    -\frac{\partial m_{\alpha, i}}{\partial t}= r_{\alpha, i} =c_{\alpha, i}\,e^{-(b_{\alpha, i}/T)}\,m_{\alpha, i},
\end{equation}
where $T$ [\text{K}] is the absolute temperature. The  pre-exponential factor, $c_{\alpha, i}$ [$\text{s}^{-1}$], and the activation temperature, $b_{\alpha, i}$ [$\text{K}$], may vary between live and dead moisture states \cite{Rossa_2017}. Nonetheless, for a specific moisture level, these parameters are taken as species-independent \cite{DiBlasi_2008}, yielding the simplifications $c_{\alpha, i} \approx c_{\alpha}$ and $b_{\alpha, i} \approx b_{\alpha}$.

The second exothermic stage, representing the combustion of volatiles and char, is described by first-order Arrhenius kinetics for $m_{\beta, i}$ and by half-order kinetics for $C_{int}$, the local oxygen concentration at the gas-solid interface, yielding the following expression for each fuel component $i$:
\begin{equation}
    \label{eq:Exothermic_Reaction}
    \begin{split}
    -\frac{\partial\,m_{\beta, i}}{\partial t} &= c_{\beta, i} \, e^{-(b_{\beta, i}/T)} \, m_{\beta, i} \, C_{int}^{1/2}
    =  k_c\left(\frac{\sigma_i m_{\beta, i}}{\rho}\right)\left(C_{bulk}-C_{int}\right)\left(\frac{\text{kg}\,\text{wood}}{\text{kg}\, \text{O}_2}\right).
    \end{split}
\end{equation}
Terms $C_{bulk}$ [$\text{kg} \, \text{m}^{-3}$] and $C_{int}$ [$\text{kg} \, \text{m}^{-3}$] denote the oxygen mass concentrations in the bulk gas phase and at the active gas–solid interface, correspondingly, while the dimensionless parameter $(\sigma_i m_{\beta, i} / \rho)$ quantifies the available gas–solid contact area for each fuel component $i$ and $\left(\text{kg}\,\text{wood} / \text{kg}\, \text{O}_2\right) \approx 0.85$. The mass-transfer coefficient, $k_c$ [$\text{m} \, \text{s}^{-1}$], is estimated by the following expression:
\begin{equation}
    \label{eq:Mass_Transfer}
    k_c = 0.012 \, \langle u\rangle^{0.8} \, L^{-0.2},
\end{equation}
based on the classical scaling of the Sherwood number ($Sh\sim Re^{0.8}$), and using representative transport properties for hot combustion gasses. The characteristic velocity, $\langle u \rangle$ [$\text{m} \, \text{s}^{-1}$], is the Euclidean norm of the ambient wind speed, $u_w$, and the maximum buoyant velocity, $u_{buoy, m}$, while the characteristic length scale $L$ [\text{m}] is the flame front depth (all quantities derived in the following subsections). The stoichiometric wood-to-oxygen mass ratio (last term in Eq.~(\ref{eq:Exothermic_Reaction})), alongside other stoichiometry-related coefficients introduced later, is estimated from the cellulose combustion reaction ($\text{C}_6\text{H}_{10}\text{O}_5 + 6\text{O}_2 \rightarrow 6\text{CO}_2 + 5\text{H}_2\text{O}$). Although wildland biomass inherently contains lignin and hemicellulose, standardizing on cellulose provides a sufficient approximation for macroscopic rate of spread calculations, as it constitutes the dominant structural component and primary energetic driver of dry plant biomass \cite{Darmawan_2016, Tuttle_1978}.

The interfacial concentration, $C_{int}$, may readily be eliminated by combining the two expressions in Eq.~(\ref{eq:Exothermic_Reaction}), which represent, respectively, the kinetic and the mass-transfer resistance to the reaction. The final result is the following expression in terms of the dimensionless ratio $\epsilon_r$:
\begin{equation}
    \label{eq:Exothermic_Final}
    \begin{split}
    -\frac{\partial\,m_{\beta, i}}{\partial t} &= r_{\beta, i} 
    = c_{\beta, i}\,e^{-(b_{\beta, i}/T)}\,m_{\beta, i}\,C_{bulk}^{1/2}
    \left(\sqrt{\epsilon_r^2+1} - \epsilon_r\right),
    \end{split}
\end{equation}
with,
\begin{equation}
    \label{eq:epsilon_r}
    \epsilon_r = \frac{\rho \, c_{\beta, i}\,e^{-(b_{\beta, i}/T)}}{1.7 \, k_c \sigma_i \sqrt{C_{bulk}}}.
\end{equation}
The activation temperature, $b_{\beta, i}$ [\text{K}], is assumed to be invariant across different vegetation species, yielding $b_{\beta, i} \approx b_{\beta}$. In contrast, the pre-exponential factor, $c_{\beta, i}$ [$\text{m}^{3/2} \, \text{kg}^{-1/2} \, \text{s}^{-1}$], is expected to depend strongly on the specific SAV, $\sigma_i$ [$\text{m}^{-1}$], of each fuel component. This dependence is quantified by resorting to the empirical correlation for the maximum reaction velocity, $\Gamma'_{max} = \big[0.0591 + 2.926(\sigma_i/100)^{-1.5}\big]^{-1}$ [$\text{min}^{-1}$], derived from Rothermel's laboratory experiments \cite{Andrews_2018}. Our proposed kinetic factor is formulated as $c_{\beta, i} = \gamma_r \, \Gamma'_{max}$, where $\gamma_r$ serves as a calibration constant.

\subsection{Thermal energy balance}
    \label{subsec:Energy_Balance}

Under the assumption of local thermal equilibrium, the energy conservation equation for the control volume $(dx,dy,H)$ is approximated as follows: 
\begin{equation}
    \label{eq:Energy_Balance}
    \begin{split}
    \left(m \, c_p + m_g \, c_{pg}\right) \frac{\partial T}{\partial t} = \sum_{i=1}^{N}\left(-h_{\alpha} r_{\alpha, i} + h_{\beta} r_{\beta, i}\right)  + m_g c_{pg} \nabla \cdot \left( \mathbf{D}_{\mathrm{eff}} \nabla T \right)  - m_g c_{pg} \nabla \cdot \left( \mathbf{u}_{\mathrm{eff}} T \right) - U \, (T-T_a).
    \end{split}
\end{equation}
Parameters $c_p$ [$\text{J} \, \text{kg}^{-1} \, \text{K}^{-1}$] and $c_{pg}$ [$\text{J} \, \text{kg}^{-1} \, \text{K}^{-1}$] denote the specific heat capacities of the solid  and gas phases, $h_\alpha$ [$\text{J} \, \text{kg}^{-1}$] and $h_\beta$ [$\text{J} \, \text{kg}^{-1}$], the specific heats of vaporization and combustion. In addition, $T_a$ [$\text{K}$] is the ambient temperature, and $\nabla$ is the gradient operator. The subsequent terms on the right hand side of Eq.~(\ref{eq:Energy_Balance}) describe respectively short-range (diffusion like) interactions, advection through the fuel bed and loss to the ambient.

The exact expressions used for the transport coefficients $\mathbf{D}_{\mathrm{eff}}$, $\mathbf{u}_{\mathrm{eff}}$, and $U$ are provided in the following subsections, after the definition of the appropriate velocity and length scales and the exposition of the procedures for estimating parameters such as the flame tilt angle, the maximum fireline intensity, and the local buoyant velocity field. Equations~(\ref{eq:Endothermic_Reaction}), (\ref{eq:Exothermic_Final}) and (\ref{eq:Energy_Balance}) are integrated in time by an explicit finite difference numerical scheme. Spatial derivatives in Eq.~(\ref{eq:Energy_Balance}) are discretized using upwind differences for the advection terms and central differences for the diffusion terms. At the computational boundaries, open-outflow conditions are prescribed to prevent artificial numerical reflections \cite{Papanastasiou_1992, Dong_2014}.

\subsection{Shear-induced flow field}
    \label{subsec:Shear_Induced_Flow}

The flow field in the fire zone induced by ambient wind shear is approximated by a combination of the flow conditions downstream of the fire zone (i.e. over intact canopy), quantified by the mean velocity $u_f$, through the canopy and upstream of the fire zone (i.e. over burn remains), quantified by the mean velocity $u_b$. The mean value in the $z-$direction over the plantation height $H$ is the shear-induced velocity contribution, $\mathbf{u}_{\textnormal{shear}} = u_{shear} \mathbf{e}_\mathbf{w}$, to the effective velocity field in Eq.~(\ref{eq:Energy_Balance}), with $\mathbf{e}_\mathbf{w}$ the unit vector in the direction of the wind. 
The flow field over intact canopy is taken to result from a balance between ambient shear and momentum dissipation induced inside the canopy \cite{Inoue_1963, Finnigan_2007}. A comprehensive derivation of the velocity $u_f$, based on the work of Harman and Finnigan \cite{Finnigan_2007}, is provided in Appendix A1 (Supplementary Material).

The upstream flow field over burned remains is expected to follow the standard logarithmic profile over rough terrain, with $u_{log}$ the mean air velocity from the ground up to the height $H$. However, not all of this airflow enters the control volumes inside the flame zone, because part of it may be deflected \cite{Vogiatzoglou_2024, Canfield_2014, Cheney_1995, Finney_2020}, moving either around or above the flame (blue vectors in Fig.~\ref{fig:Figure_2}). The net air intake into the control volume in the reactive zone, denoted by the mean inflow vector, $\mathbf{u}_{\mathbf{b}} = u_b \mathbf{e}_\mathbf{w}$ is a key variable, as it serves the dual purpose of supplying combustion oxygen while simultaneously modulating flame temperature through convective cooling \cite{Pastor_2003}. Thus, estimation of $u_b$ remains an important task in wildland fire modeling \cite{Albini_1981, Beer_1991, Nelson_2012}.

Within the ADfiRe framework, $u_b$ is calculated by the following expression:
\begin{equation}
    \label{eq:Downstream_Velocity}
    u_b = \delta\, u_{buoy, m} + u_{log}\left(1-e^{-\gamma w}\right)\,\eta_b.
\end{equation}
The first term represents air intake by pure entrainment of the buoyant plume. This term is important only in the total absence of wind, as it is known that the introduction of even minor wind effects  disrupts the symmetric entrainment field, shifting the dominant air influx to the rear margins of the fire zone \cite{Beer_1991}. The second term estimates the reduction of the mean inflow velocity, $u_b$, from the mean over the burned terrain, $u_{log}$, which is caused by deflection around the flame of transverse width $w$ [\text{m}] and by deflection above the control volume (due to parameter $\eta_b$).

Parameter $\eta_b$ quantifies the decreasing flow resistance to the deflection of airflow above the fire zone as the flame becomes gradually more tilted. This difference in behavior between a plume- and a wind-dominated flame is well-documented in high-fidelity CFD simulations \cite{Finney_2015, Finney_2022}, and is introduced by the following semi-empirical relation based on Byram number $N_c$:
\begin{equation}
    \label{eq:Entrainment_Coefficient}
    \eta_b = 0.4 \left[\tanh \left(\frac{N_c}{2} - 3 \right) + 1.5 \right].
\end{equation}
Accordingly, $\eta_b\approx 1$ for $N_c > 10$ (plume-driven) and $\eta_b\approx 0.2$ for $N_c < 2$ (wind-driven). The asymptotic value $\eta_b\approx 1$ for $N_c > 10$ is consistent with the recommendation of Nelson et al. \cite{Nelson_2012}. However, those authors proposed a constant value of $\eta_b$ = 0.71 for all $N_c < 10$, whereas the present model presumes a stronger decrease of $\eta_b$ with decreasing $N_c$.

Finally, the flow field inside the fire zone that is induced by wind shear is approximated by interpolating between the (extreme) mean velocities $u_f$ in the front and $u_b$ in the back of the fire zone, according to the expression:
\begin{equation}
    \label{eq:Shear_Velocity}
    u_{shear}=u_f \left(\frac{\mathrm{LAI}}{\mathrm{LAI}_0}\right) + u_b \left(1-\frac{\mathrm{LAI}}{\mathrm{LAI}_0}\right).
\end{equation}
 The interpolation is dynamically weighted by the instantaneous leaf-area-index, $\text{LAI} = (1 / \rho) \sum_{i=1}^{N} m_{i} \sigma_i$, relative to the initial condition, $\text{LAI}_0$, reflecting the progressive attenuation of canopy-induced drag as fuel mass is consumed.

\subsection{Dynamics of flame inclination}
    \label{subsec:Flame_Tilting}

\begin{figure}
    \centering
    \includegraphics[width=.4\textwidth]{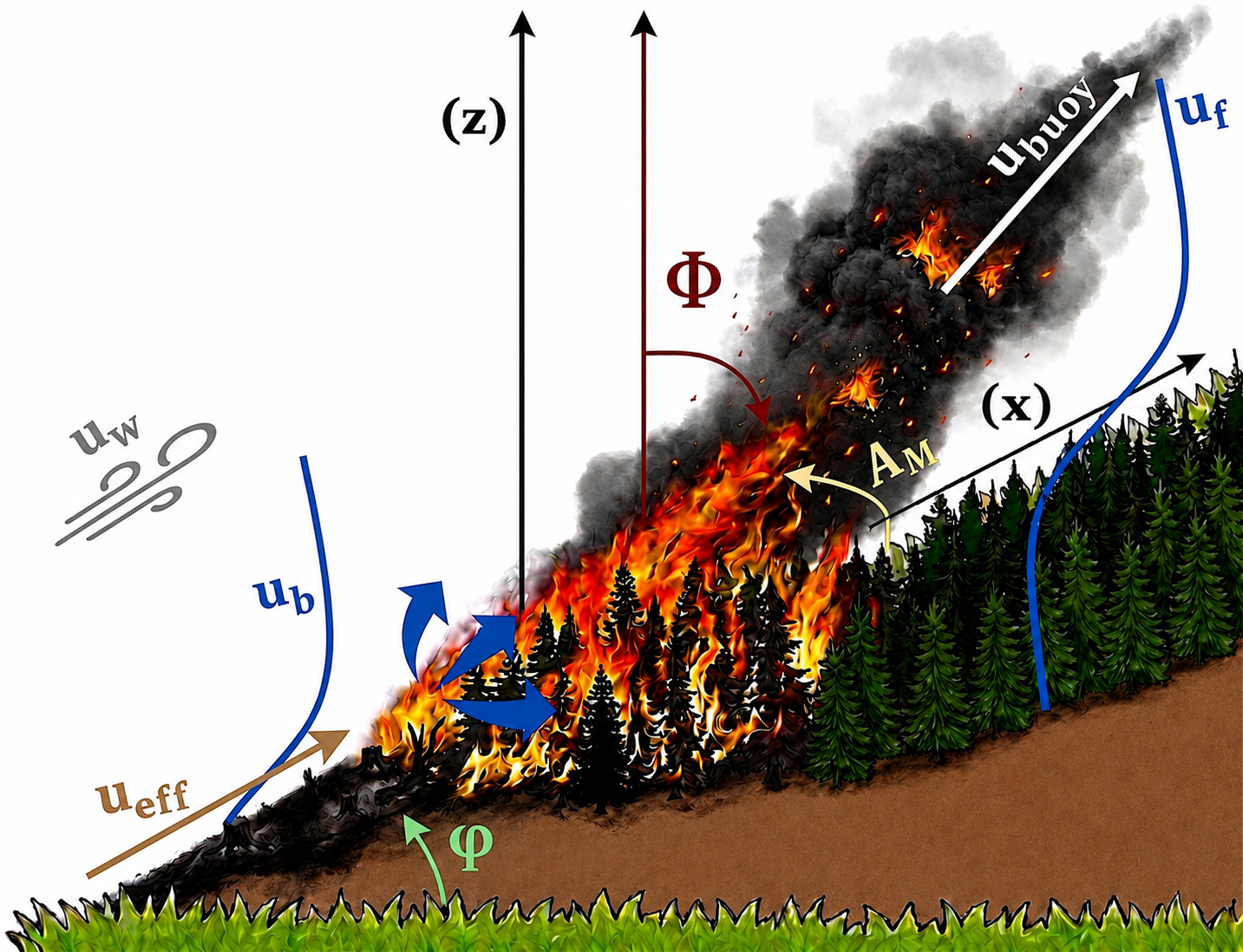}
    \caption{Geometric formulation of 1D fire spread across an inclined topography. The figure defines the local terrain inclination ($\varphi$), and the flame angle deviation from the vertical ($\Phi$), with the resultant velocity vectors.}
    \label{fig:Figure_2}
\end{figure}

Building upon the foundational approach of Nelson \cite{Nelson_2002}, the ADfiRe framework incorporates the effects of local topography and wind-driven flame deflection into a mean velocity $\mathbf{u}_{\mathrm{proj}}$ that augments the shear-induced velocity $\mathbf{u}_{\mathrm{shear}}$ through the canopy. The methodology is sketched in Fig.~\ref{fig:Figure_2} for 1D fire propagation in the direction of the wind, over a terrain with uniform slope, $\varphi$. If the tilt angle of the flame due to the wind is $\Phi$, then the projection of the buoyant velocity distribution, $u_{buoy}(x)$, on the ground is $\big(u_{buoy}(x)\sin(\Phi+\varphi)\big)$ and the additional contribution to the mean advection velocity through the canopy is estimated as:
\begin{equation}
    \label{eq:Projection_Velocity_1D}   \mathbf{u}_{\textnormal{proj}}=\left(\frac{u_{log,H}}{u_w}\right)u_{buoy}(x) \sin(\Phi + \varphi) \, \mathbf{e}_\mathbf{x}.
\end{equation}
Term ($u_{log, H} / u_w$) is the ratio of velocities at heights $z=H$ and $z=$ 10 \text{m} in the logarithmic profile, and serves as a scaling factor for the effect of buoyancy on the control volume. Indeed, while $u_w$ and $u_{buoy, m}$ are characteristic scales of the atmospheric boundary layer and the macroscopic buoyant plume, the energy balance, Eq.~(\ref{eq:Energy_Balance}), is applied only up to the height $H$.

\begin{figure}
    \centering
    \includegraphics[width=.6\textwidth]{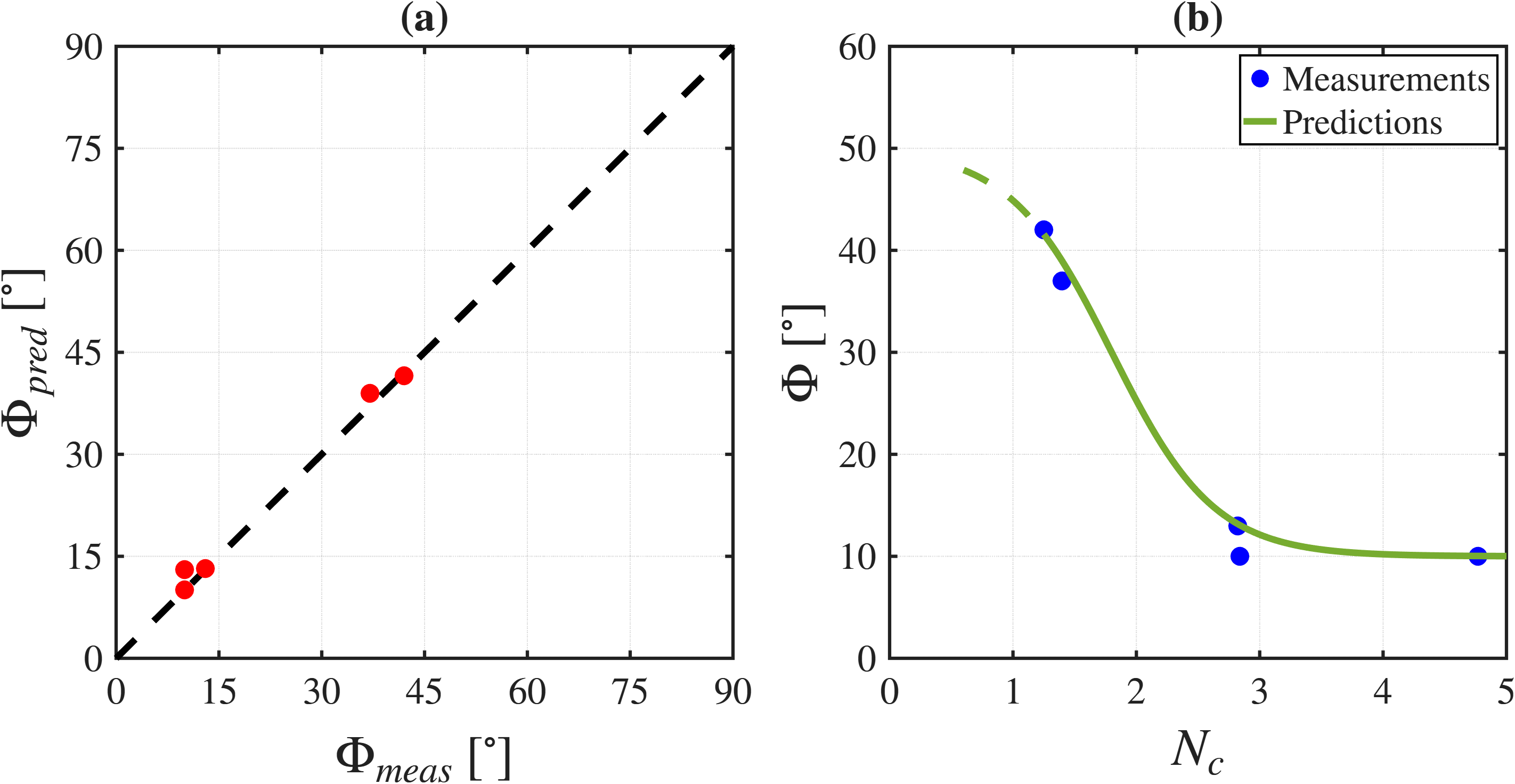}
    \caption{(a) Parity plot comparing experimentally measured flame deflection angles ($\Phi_{meas}$) and theoretical predictions ($\Phi_{pred}$) derived from Eq.~(\ref{eq:Tilt_Angle}). (b) Functional dependence of flame tilt angle ($\Phi$) on the dimensionless Byram number ($N_c$), comparing the empirical dataset (blue dots) \citep{Morandini_2010} against the continuous model formulation of Eq.~(\ref{eq:Tilt_Angle}) (green curve).}
    \label{fig:Figure_3}
\end{figure}

A correlation is developed for the flame tilt angle due to the wind, based on the data provided by Morandini and Silvani \cite{Morandini_2010} and the analysis of Nelson \cite{Nelson_2015}. The deviation angle, $\Phi$, from the vertical is a function of the Byram convection number, $N_c$, which compares wind inertia to fire-induced buoyancy, thereby governing the transition between wind- and plume-dominated  regimes \cite{Morvan_2018, Trouve_2020, Morvan_2025}. The Byram number is currently defined as \cite{Nelson_2015, Morvan_2025}:
\begin{equation}
    \label{eq:Byram_Number}
    N_c = \left(\frac{u_{buoy, m} \cos\varphi}{u_w - \text{ROS} + u_{buoy, m} \sin\varphi}\right)^3
\end{equation}
in terms of the maximum, $u_{buoy,m}$, of the local distribution of buoyant velocity and the tilt angle, $\Phi$, is estimated as:
\begin{equation}
    \label{eq:Tilt_Angle}
    \Phi(N_c) = 30 - 20\tanh\big[1.2 \, (N_c - 1.8)\big].
\end{equation}
The performance of Eq.~(\ref{eq:Tilt_Angle}) is depicted in Figure~\ref{fig:Figure_3}(a), which compares predictions and data for all five experiments of Morandini and Silvani \cite{Morandini_2010}, and in Figure~\ref{fig:Figure_3}(b) which illustrates the typical dependence of $\Phi$ on $N_c$. It is noted that the functional relation between $\Phi$ and $N_c$ in Eq.~(\ref{eq:Tilt_Angle}) is motivated by the agreement with the available data (five points), which indicate a rather sharp increase in $\Phi$ with a decrease of $N_c$ below the value $N_c = 3$. The behavior for $N_c < 1$ is uncertain, though a reasonable limit for $N_c \rightarrow 0$ would be $\Phi \rightarrow 90^o - \phi$, i.e. a flame tangent to the ground.

\subsection{Fireline evolution and estimation of buoyancy}
    \label{subsec:Evolution_Buoyancy}

The location of the fireline is identified by scanning the temperature field for local maxima. Specifically, for each fixed lateral coordinate $y_k$ (where $k = 1, 2, \dots, N_y$), the temperature field is sampled along the $x$-direction to isolate the local peak, $T_{max}^{(k)}$, and its associated coordinates $(x_k, y_k)$. The global peak, $T_{max}(t)$, located at the grid point $l_1 = (x_{max}, y_{max})_t$, serves as the reference point for the fire front at time $t$. The procedure is repeated after an interval $\Delta t$ [\text{s}] (a multiple of the fundamental integration step $d_t$ [\text{s}]) and identifies the updated global peak location at $l_2 = (x_{max}, y_{max})_{t+\Delta t}$. The vector $\mathbf{l}/\Delta t = (l_2 - l_1)/\Delta t$ gives the direction and speed of fire propagation.

The evolution of the remaining fuel load, $m_{\beta}$, over the interval $\Delta t$ allows for the estimation of the peak fireline intensity, $I_{B, max}$ [$\text{W} \, \text{m}^{-1}$], by the relation:
\begin{equation}
    \label{eq:Fireline_Intensity}
    I_{B, max} = \frac{1}{\Delta t} \int_{l_1-e}^{l_2+e}{\left[m_{\beta}(t)-m_{\beta}(t+\Delta t)\right] h_{\beta}\,dl}.
\end{equation}
The integral is evaluated numerically using a trapezoidal scheme based on the linear interpolation of the discrete nodal mass values, with the limits of integration extended by an arbitrary length $e$ in both directions to capture the full spatial extent of the combustion zone.

The maximum fireline intensity, $I_{B,  max}$, is subsequently employed to determine the maximum buoyancy-induced velocity, $u_{buoy, max}$, via the well-known formula:
\begin{equation}
    \label{eq:Maximum_Buoyancy}
    u_{buoy, max}=\left(\frac{2 \, g \, I_{B, max}}{\rho_{g,0} \, c_{pg,0} \, T_a}\right)^{1/3},
\end{equation} 
where, $\rho_{g,0}$ [$\text{kg} \, \text{m}^{-3}$] and $c_{pg,0}$ [$\text{J} \, \text{kg}^{-1} \, \text{K}^{-1}$] denote the air density and specific heat capacity at ambient temperature $T_a$. Using $u_{buoy, max}$ as a scale, the buoyant velocity profile is approximated as follows in terms of the temperature distribution $T(x, y, t)$:
\begin{equation}
    \label{eq:Buoyant_Velocity}
    u_{buoy}(x, y)=u_{buoy, max}\left(\frac{T(x, y)-T_a}{T_{max}(t) - T_a}\right).
\end{equation}
In particular, the local maxima $u_{buoy, m}$ of the buoyant velocity along the entire flame front, which are required for the evaluation of the local Byram number in Eq.~(\ref{eq:Byram_Number}), are given by:
\begin{equation}
    \label{eq:Maximum_Buoyant_Velocity}
    u_{buoy, m}=u_{buoy, max}\left(\frac{T_{max}^{(k)} - T_a}{T_{max}(t) - T_a}\right).
\end{equation}
Equations~(\ref{eq:Buoyant_Velocity}) and (\ref{eq:Maximum_Buoyant_Velocity}) rely on the commonly adopted assumption that the self-similar velocity and temperature profiles exhibit the same functional shape and can, therefore, be expressed as normalized ratios with respect to their corresponding local maximum values \cite{Turner_1979, George_1994}. 

\subsection{Computation of characteristic lengthscales and transport properties}
    \label{subsec:Lengths_Diff_U}

The current analysis further enables the calculation of two characteristic length scales defining the flame geometry: (i) the flame zone depth, $L$ [\text{m}], at the fire peak, and (ii) the fire front width, $w$ [\text{m}], measured orthogonal to the ambient wind vector. The depth, $L$, is defined as the distance along the propagation vector $\mathbf{l}$ from the global peak, $T_{\textnormal{max}}(t)$, to the isothermal contour defined by a threshold temperature given by $T_L = T_a + 0.35 (T_{max}(t) - T_a)$. The fire front width, $w$, is established by analyzing the array of local temperature maxima, $T_{max}^{(x_i,y_i)}$, and defining the endpoints $A = (x_A, y_A)$ and $B = (x_B, y_B)$ as the locations where the local peak temperature decays to $T_{max}^{(x_i,y_i)} \approx 550 \,$ \text{K}. The width $w$ is then determined by the projection of the segment $\overline{AB}$ onto the axis normal to the ambient wind direction.

Based on the above scales, the effective diffusivity, $\mathbf{D}_{\textnormal{eff}}$, that appears in Eq.~(\ref{eq:Energy_Balance}) is expressed as follows:
\begin{equation}
    \label{eq:Diffusion_Coefficient}
    \mathbf{D}_{\mathrm{eff}} = A_{b}\,u_{buoy}\,L + A_{d} \, \mathbf{u}_{\mathrm{shear}}\,L + \frac{k_r}{\rho_g\,c_{pg}}.
\end{equation}
Equation~(\ref{eq:Diffusion_Coefficient}) includes dispersion induced by the buoyant plume and by wind-shear, both modeled as proportional to the product of a local characteristic velocity and a macroscopic length scale \cite{Cussler_2009}. The final term accounts for radiative exchange in the dense-medium limit, expressed as $k_{r} = 5.33 \sigma_b T^3 / k_d$ \cite{Siegel_1972}, where $\sigma_b$ is the Stefan-Boltzmann constant and $k_d$ [$\text{m}^{-1}$] is the extinction coefficient of the medium, as measured for combustion of forest fuels by Agueda et al \cite{Agueda_2010}.

The treatment of the combustion mixture as an optically dense medium (an assumption that greatly simplifies computations of radiative heat transfer) is supported by the existence in the flame of high concentrations of soot and other particulate matter \cite{Howell_2010, Mossi_2012}. The optical thickness of the flame, defined as the ratio of the characteristic length of the problem to the extinction length, is a-posteriori calculated in the range $2.5<L/(1/k_d)<7.5$, which is considered satisfactory in the context of the present approximate methodology.

The loss coefficient, $U$, from the top of the control volume to the ambient is expresses as follows:
\begin{equation}
    \label{eq:Total_Convection_Coefficient}
    \begin{split}
    U &= \big(h_{nc}^3 + h_{fc}^3\big)^{1/3} + \varepsilon\sigma_{b} \left(T^2 + T_a^2\right)\big(T + T_a\big) + \rho_g c_{pg} \left(\frac{u_{log, H}}{u_w}\right) u_{buoy}\cos\big(\Phi+\varphi\big).
    \end{split}
\end{equation}
The three contributions in Eq.~(\ref{eq:Total_Convection_Coefficient}) represent, respectively: (i) mixed convection losses to the ambient (estimated from standard correlations for a hot horizontal surface \cite{Bergman_2011, Churchill_1983}) as outlined in Appendix A3 (Supplementary Material) (ii) radiative losses to the ambient (in terms of the emissivity $\varepsilon$ of the upper surface of the burning canopy), and (iii) sensible losses by the flue gases exiting from the top of the control volume. This last contribution is estimated from the component of buoyant velocity normal to the ground, scaled down by the ratio ($u_{log,H}/u_w$) that was also used in the definition of $\mathbf{u}_{\mathrm{proj}}$.

The last parameter required for the closure of the model is the bulk oxygen concentration, $C_{bulk}$, that appears in Eq.~(\ref{eq:Exothermic_Final}) of the kinetic model. Due to the intense mixing in the flame zone \cite{Sun_2026}, $C_{bulk}$ is taken as uniform through the entire flame depth at a given location of the fire front, but may vary with the location along the front and also with time.  As the oxygen concentration inside the flame is below the value $C_{air}$ for ambient fresh air due to dilution by gaseous combustion products, $C_{bulk}$ is modeled as follows:
\begin{equation}
    \label{eq:Oxygen_Bulk_Concentrantion}
    C_{bulk} = C_{air} \, \frac{F_{air}}{F_{air} + F_{flue}}.
\end{equation}
Term $F_{air}$ [$\text{m}^2 \, \text{s}^{-1}$] denotes the volumetric flow rate of fresh air entering the flame zone  and $F_{flue}$ [$\text{m}^2 \, \text{s}^{-1}$] denotes the volumetric production rate of combustion flue gases. Utilizing cellulose stoichiometry and ideal gas transformations evaluated at the local mean temperature, $T_{mean}^{(k)} = (T_{max}^{(k)} + T_a) / 2$, $F_{flue}$ term is computed as follows:
\begin{equation}
    \label{eq:Flue_Gases}
    \begin{split}
    F_{flue} &= \left(0.0046 \, T_{mean}^{(k)} \right) \sum_{i=1}^{N} \int_{(x_k - L, y_k)}^{(x_k + L, y_k)}r_{\alpha, i} \, dx + \left(0.0056 \, T_{mean}^{(k)} \right)\sum_{i=1}^{N}\int_{(x_k - L, y_k)}^{(x_k + L, y_k)}r_{\beta, i} \,dx.
    \end{split}
\end{equation}
The air inflow, $F_{air}$, is estimated in terms of the mean air velocity, $u_b$, entering the control volume by the expression: 
\begin{equation}
    \label{eq:Air_Flow}
    F_{air} = u_b \, H \left(1 + \frac{0.04}{H^3}\right).
\end{equation}
Equation~(\ref{eq:Air_Flow}) gives the expected result $F_{air}\approx u_b H$ for all canopy heights except for very short ones. When the canopy is very short, turbulent mixing penetrates through the control volume (of height $H$) and $C_{bulk}$ gradually approaches $C_{air}$. This behavior is modeled by the last term inside the parenthesis of Eq.~(\ref{eq:Air_Flow}).

Finally we discuss the modification of the temperature field and  compare with experimental data. Similar to the argument in Eq.~\ref{eq:Projection_Velocity_1D}, the control volume of the model is restricted to a height $H$, whereas the measured temperature is affected by the entire thermal plume, which entrains ambient air of temperature $T_a$. Thus, the same scaling factor, ($u_{log, H} / u_w$), is used to define the scaled temperature $T_{sc}$, corresponding to the entire plume, as follows:
\begin{equation}
    \label{eq:Scaling_Temperature}
    T_{sc} = T\left(\frac{u_{log,H}}{u_w}\right) + T_a\left[1 - \left(\frac{u_{log,H}}{u_w}\right)\right].
\end{equation}

\section{Results}
    \label{sec:Results}

\subsection{The validation data and the model parameters}
    \label{subsec:Data}

We validate the present model on data from  five field campaigns by Morandini and Silvani \cite{Morandini_2010}, performed under well-controlled conditions and exhibiting 1D fire propagation. These experiments were conducted in living shrubland plots in southern France, dominated by Syrmatium Glabrum (SG), Arbutus Unedo (AU), Genista Lobelia (GL), Erica Arborea (EA), and Asphodelus Albus (AA). Based on temperature and heat flux measurements, this study classified propagation into wind-driven, plume-driven and intermediate regimes, questioning the conventional Byram number limits ($N_c < 2$ and $N_c > 10$). Nelson \cite{Nelson_2015} showed that reasonable $N_c$ values emerge only when wind speed is augmented by the ground-parallel component of the buoyant velocity.

The composition of the canopy for each of the five field experiments is as follows:  Experiments 1 and 3 ($\text{E}_1$ and $\text{E}_3$) were pure stands, SG and GL respectively, while $\text{E}_2$, $\text{E}_4$ and $\text{E}_5$ featured mixtures of AU, GL, EA, and AA. Following standard practice \cite{Mell_2007, Sullivan_2009, Pigg_2024}, and in the absence of high-resolution species-distribution data, spatially averaged properties were assigned to the mixed plots, with effective SAVs: $\sigma_{E_1}$ = 8000 m$^{-1}$, $\sigma_{E_2}$ = 7000 m$^{-1}$, $\sigma_{E_3}$ = 9880 m$^{-1}$, and $\sigma_{E_4} = \sigma_{E_5}$ = 7600 m$^{-1}$.

The above values are  engineering estimates, as it is well-known that the SAV value may vary widely with the season of the year and the relative content of live and dead leaves, twigs, and branches of different sizes \citep{Essaghi_2017}. We note that ADfiRe permits the exploitation of more detailed data by modeling the combustion of multiple fuels with different SAV values. The trends from an indicative computation with two fuels, representing schematically fines and branches, are discussed later.

The values of all model parameters are shown in Table~\ref{table:Table_1}. The first column contains the physical parameters, whose values are more or less standard in the literature. The lower half of the second column lists the adjustable parameters of the model, which are held constant in predicting the five field datasets of Morandini and Silvani.

\begin{table}
    \centering
    \caption{Fixed thermophysical parameters used across all simulation scenarios.}
    \label{table:Table_1}
    \renewcommand{\arraystretch}{1.5}
    \begin{tabular}{ccc|ccc}
    \toprule
    \textbf{Parameter} & \textbf{Value} & \textbf{Unit} & \textbf{Parameter} & \textbf{Value} & \textbf{Unit} \\
    \midrule
    $\rho_{g,0}$ & 1.2 & $\text{kg m}^{-3}$ & $c_a$ & 30.0 & $\text{s}^{-1}$ \\
    $c_{pg,0}$ & 1000.0 & $\text{J kg}^{-1} \, \text{K}^{-1}$ & $b_\alpha$ & 4500.0 & $\text{K}$ \\
    $\rho_g$ & 0.41 & $\text{kg m}^{-3}$ & $b_\beta$ & 7000.0 & $\text{K}$ \\
    $c_{p,g}$ & 1090.0 & $\text{J kg}^{-1} \, \text{K}^{-1}$ & $\epsilon$ & 0.9 & -- \\
    $\rho$ & 513.0 & $\text{kg m}^{-3}$ & $\gamma$ & 0.87 & -- \\
    $c_p$ & 1500.0 & $\text{J kg}^{-1} \, \text{K}^{-1}$ & $\gamma_r$ & 540.0 & -- \\
    $h_\alpha$ & $22 \times 10^5$ & $\text{J kg}^{-1}$ & $A_b$ & 0.005 & -- \\
    $h_\beta$ & $18 \times 10^6$ & $\text{J kg}^{-1}$ & $A_d$ & 0.05 & -- \\
    $g$ & 9.81 & $\text{m s}^{-2}$ & $A_{nc}$ & 0.9 & -- \\
    $\sigma_b$ & $5.67 \times 10^{-8}$ & $\text{W m}^{-2} \, \text{K}^{-4}$ & $A_{fc}$ & 7.0 & -- \\
    $\kappa$ & 0.41 & -- & $\delta$ & 0.2 & -- \\
    $C_{air}$ & 0.27 & $\text{kg m}^{-3}$ & $k_d$ & 0.60 & $\text{m}^{-1}$ \\
    \bottomrule
    \end{tabular}
\end{table}

The simulations were performed with the 1D version of the code, because the validation data are roughly uniform in the cross-stream direction. The computational domain has a length of 200 \text{m} in the direction of the wind, and it is discretized by $d_x$ = 0.1 \text{m} while the fire is ignited 50 \text{m} downstream from the wind inlet. The total duration of the simulations was 400 \text{s} and the integrations were performed with a temporal step of $d_t$ = 0.001 \text{s}. Each simulation was executed in under two minutes, using a single CPU core on a laptop (AMD Ryzen 7 5800H, 3.20 GHz, 16 GB RAM), demonstrating the computational efficiency of the algorithm and its suitability for time-constrained operational strategies.

The spatial discretization used in the reported runs is selected based on a grid convergence (sensitivity) study, whose outcome is shown in the Appendix A5 (Supplementary Material). This study indicates that the computed variables do not change significantly with further reduction of $d_x$. Also tabulated in the Appendix A5 is the percentage error for coarser discretizations and the \text{CPU} execution times for all the tested values of $d_x$.

\subsection{Predictions versus data}
    \label{subsec:Fire_Metrics}

\begin{figure}
    \centering
    \includegraphics[width=.9\textwidth]{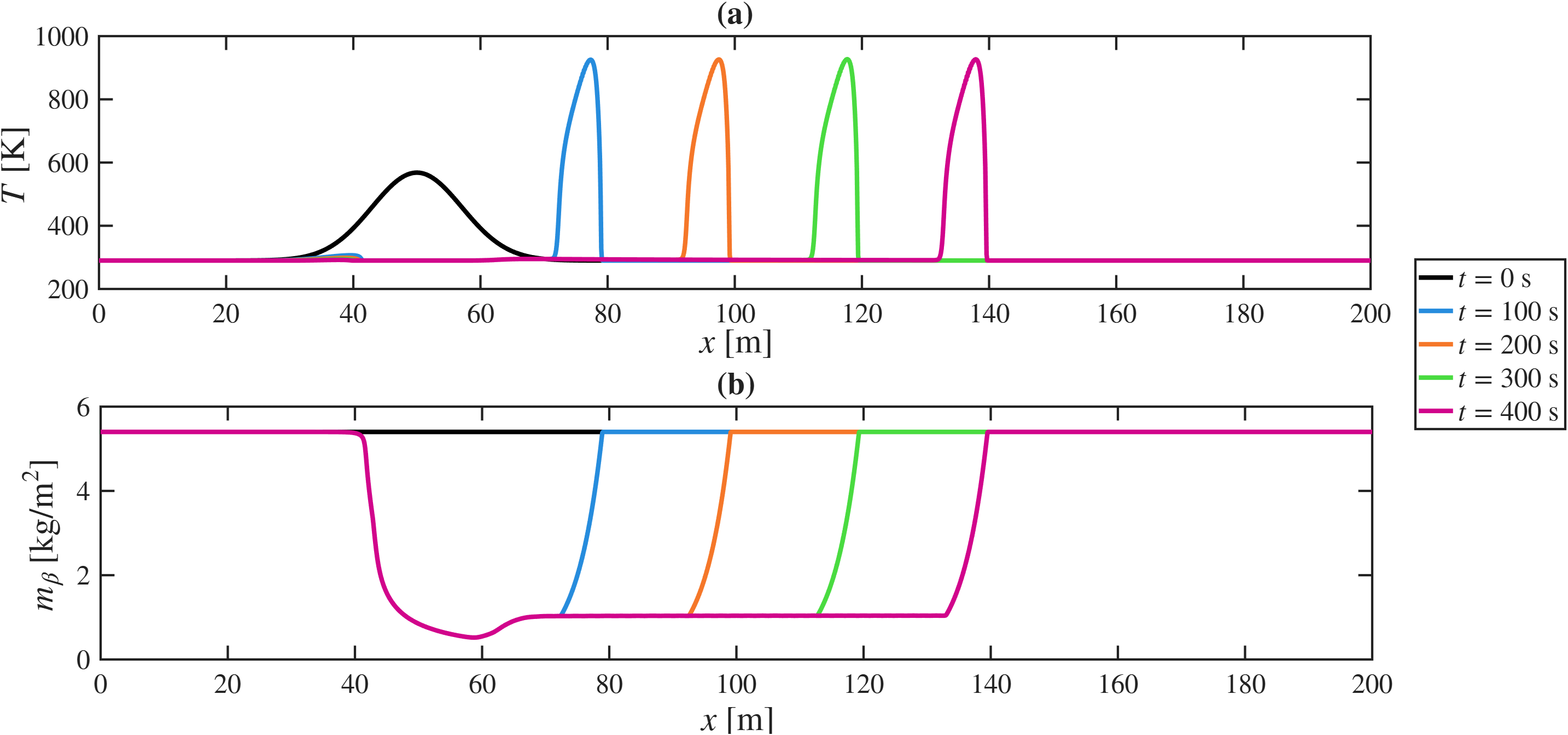}
    \caption{(a) Spatial temperature distribution, $T(x,t)$, along the $x$-direction at 100 \text{s} intervals for the $\text{E}_1$ simulation. (b) Corresponding distribution of the remaining combustible fuel load, $m_\beta(x,t)$.}
    \label{fig:Figure_4}
\end{figure}

An indicative prediction of the spatiotemporal evolution of the temperature, $T(x,t)$, and the remaining combustible mass, $m_\beta(x,t)$, in the field is depicted respectively in Figures~\ref{fig:Figure_4}(a) and \ref{fig:Figure_4}(b) and shows the rapid emergence of a fully developed and steadily progressing flame front based on the simulation of $\text{E}_1$. Predicted values of fire metrics corresponding to the fully developed region are compared to data available from the experiments in the parity plots in Figure~\ref{fig:Figure_5a} as well as in Appendix A4 (Supplementary Material).

\begin{figure}
    \centering
    \includegraphics[width=.9\textwidth]{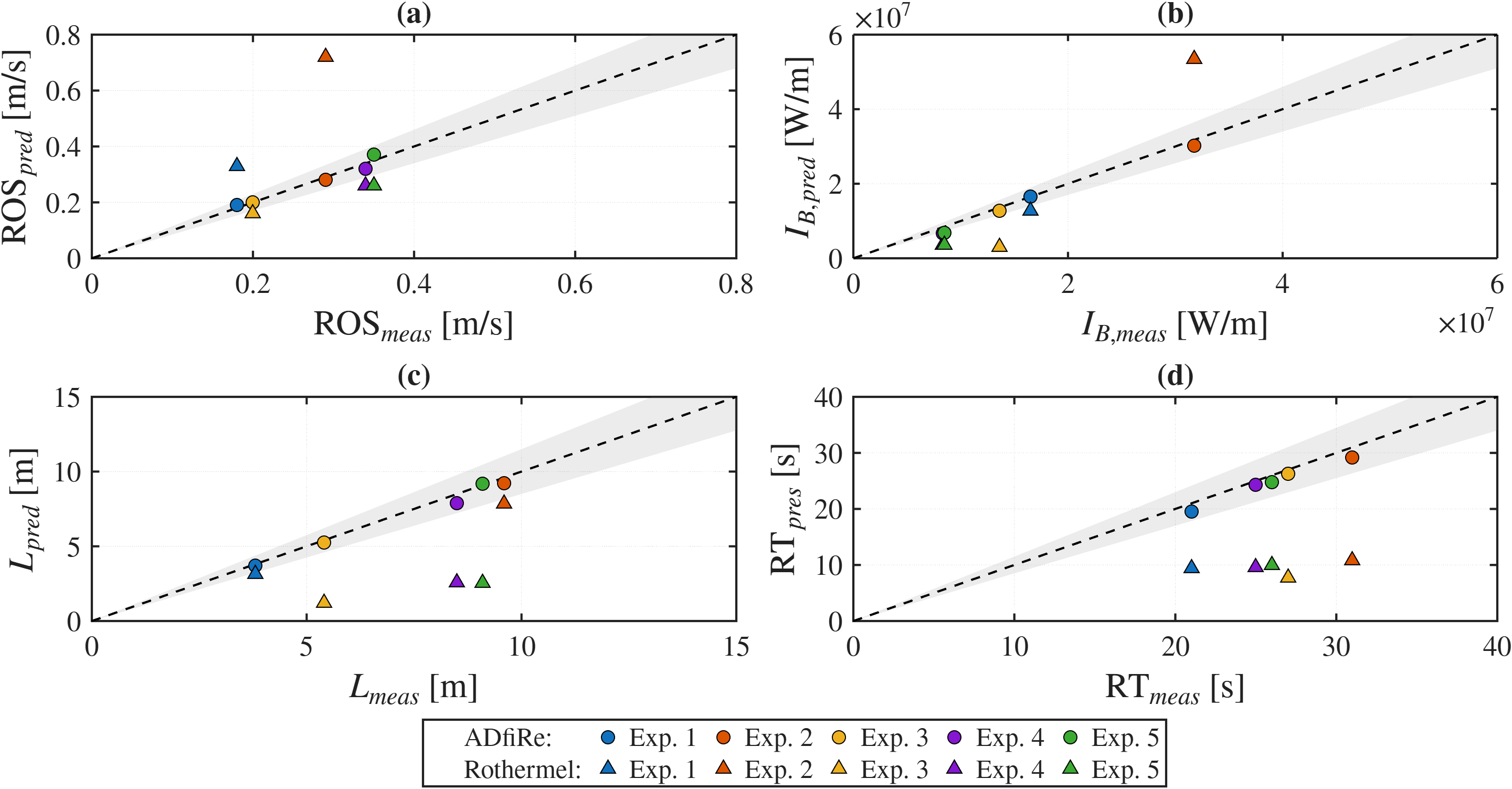}
    \caption{Comparison of ADfiRe and Rothermel models against empirical fire metrics from \cite{Nelson_2015, Morandini_2010}. Parity plots compare predicted vs. measured: (a) rate of spread - ROS, (b) fireline intensity - $I_B$, (c) fireline depth - $L$, (d) residence time - RT. Dashed line indicates 1:1 agreement, and shaded area is $\pm 15\%$ tolerance. Colors denote campaigns (Exp. 1–5).}
    \label{fig:Figure_5a}
\end{figure}

Figures~\ref{fig:Figure_5a}(a-d) refer respectively to the rate of spread, fireline intensity, fireline depth and front residence time. These parity plots compare predictions by the present model (circles) and by the standard Rothermel operational model \cite{Andrews_2018} (triangles) versus the data of Moradini and Silvani \cite{Morandini_2010}. It is evident that ADfiRe improves drastically the prediction of all parameters. Additionally, evaluations of complementary quantities of interest are provided in Appendix A4 (Supplementary Material) to further demonstrate ADfiRe's predictive accuracy.

The rate of spread (ROS) [$\text{m} \, \text{s}^{-1}$] and the fireline intensity ($I_B$) [$\text{W} \, \text{m}^{-1}$] (figures~\ref{fig:Figure_5a}a and b respectively)  are probably the most important variables for assessing the fire threat and the very good agreement of predictions by the present model with the data is particularly encouraging. To quantify the differences, the root mean square percentage error: 
\begin{equation}
    \label{eq:RMSPE}
    \begin{split}
    \text{RMSPE} &= \sqrt{\frac{1}{5}\sum_{i=1}^{5}\left(\frac{X_{meas,i}-X_{pred,i}}{X_{meas,i}}\right)^{2}},
    \end{split}
\end{equation}
was evaluated as $5\%$ for the ROS, $13\%$ for $I_B$, $12\%$ for $L$, and $9\%$ for $\text{RT}$. Even the slight underprediction of $I_B$ may be explained by the fact that Morandini and Silvani measured the remaining combustible mass roughly an hour after the passage of the fire front, so the data also include mass consumed by charring/smoldering.

\subsection{Extensions and limitations of the model}
    \label{subsec:Temperature_Profile}

\begin{figure}
    \centering
    \includegraphics[width=0.45\textwidth]{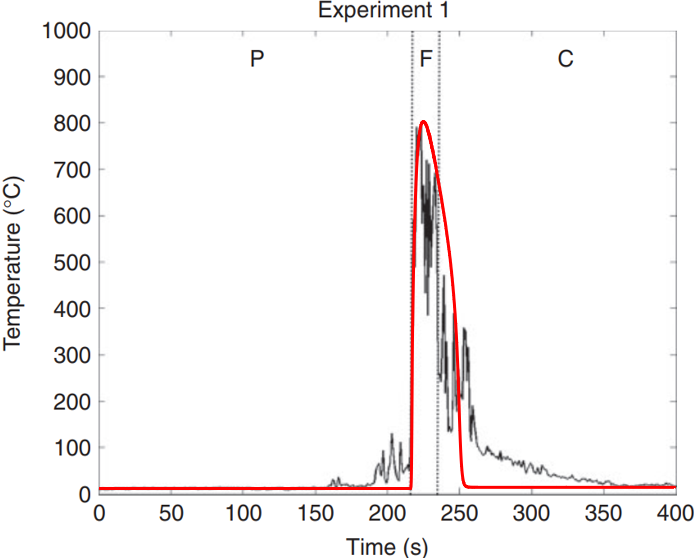}\hfill
    \includegraphics[width=0.45\textwidth]{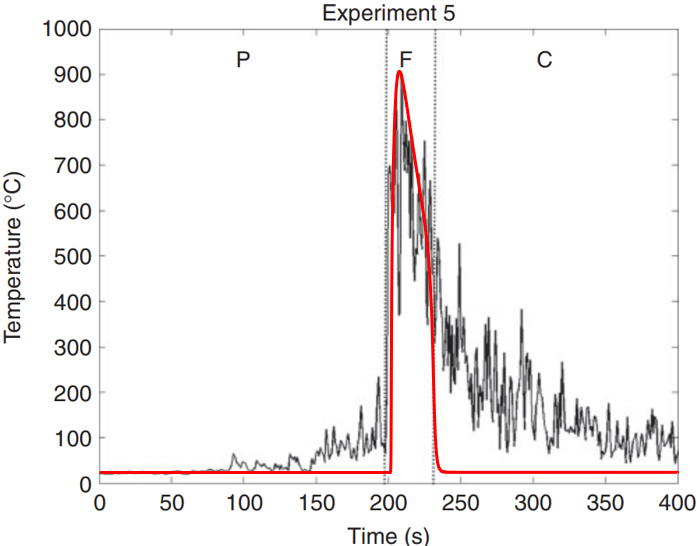}
    \caption{Comparison of the temperature distributions for $\text{E}_1$ and $\text{E}_5$. The predictions of the ADfiRe simulator (red curves) are superimposed directly onto the original experimental figures from Morandini and Silvani \cite{Morandini_2010} (black curves). The distinct preheating (P), flaming (F), and charring (C) combustion zones are clearly delineated.}
    \label{fig:Figure_6}
\end{figure}

The predicted temperature profiles for $\text{E}_1$ and $\text{E}_5$ (representative, respectively, of plume- and wind-dominated fires) are superimposed on the measurements of Morandini and Silvani \citep{Morandini_2010} in Figure~\ref{fig:Figure_6}. These predictions, which show the temperatures scaled according to Eq.~(\ref{eq:Scaling_Temperature}), provide further insights into the characteristic of the model. First, it is evident that the flaming regime is described very satisfactorily, both in terms of the residence time of the front and in terms of the maximum temperatures achieved. It is equally evident, however, that the charring regime is not adequately predicted. This failure is expected, given that charring is the result of the slow burning of larger branches that have been internally heated during the passage of the front and continue smoldering while the air has cooled. Indeed, the charring regime is characterized by a thermal decoupling between the solid and gas phases. The present single-temperature formulation assumes local thermal equilibrium and is therefore not designed to resolve these longer timescales.

\begin{figure}
    \centering
    \includegraphics[width=.49\textwidth]{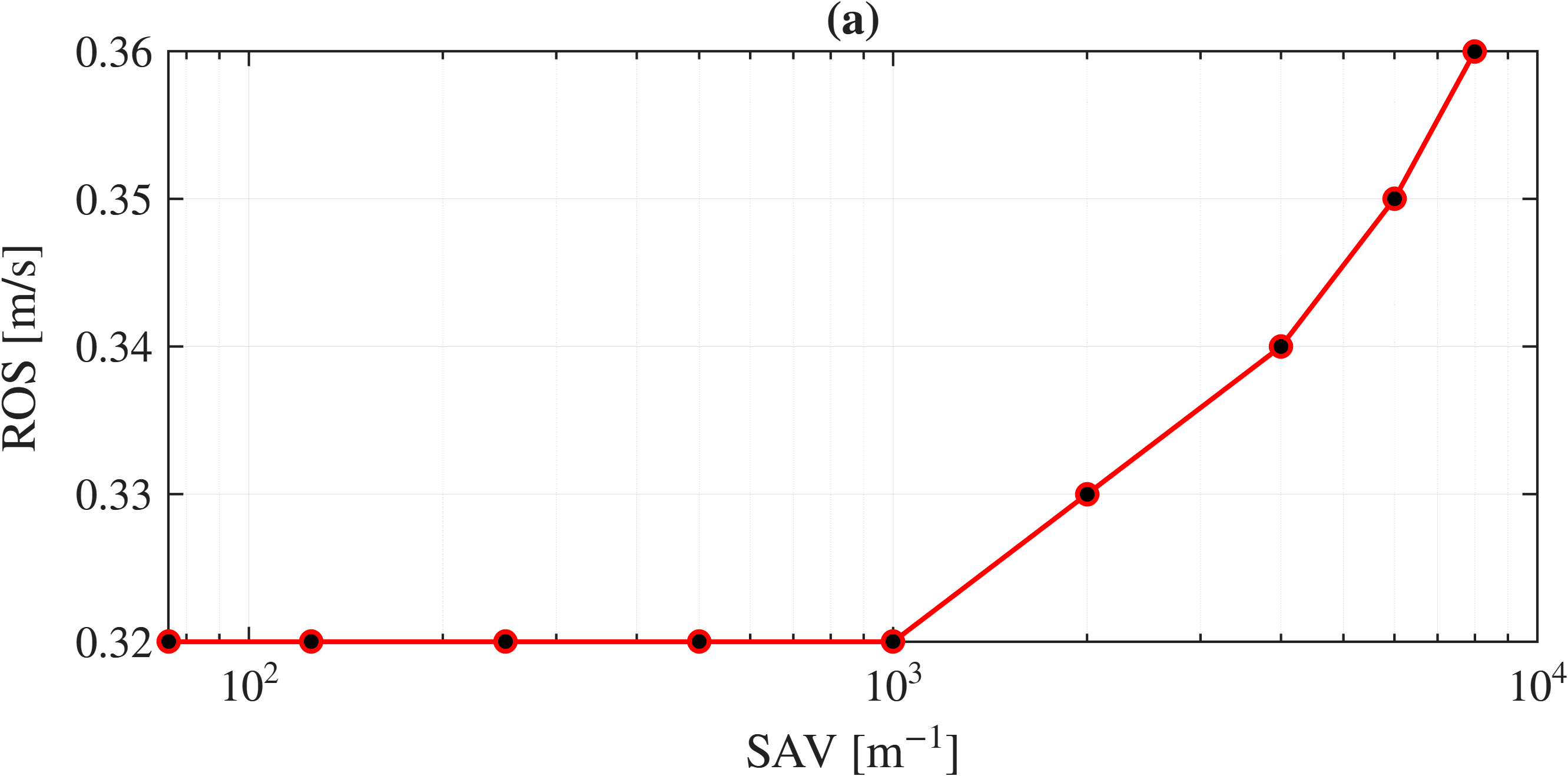}
    \includegraphics[width=.49\textwidth]{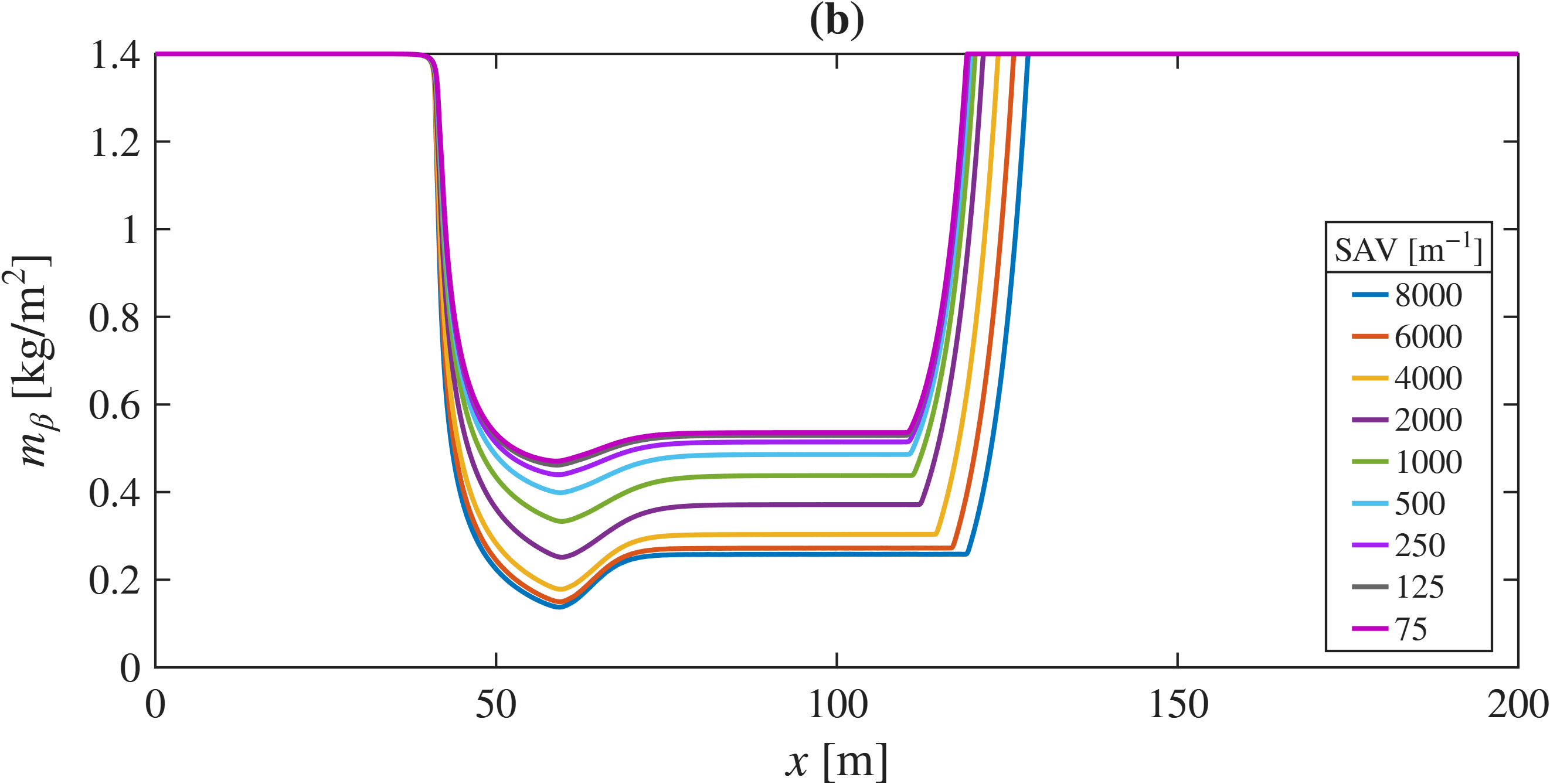}
    \caption{(a) ROS as a function of the SAV ratio for a $70/30$ fine-to-coarse fuel mixture based on the $\text{E}_5$ simulation. (b) Corresponding profile of the remaining combustible fuel load, $m_\beta$, across the domain at $t=$ 200 \text{s}.}
    \label{fig:Figure_7}
\end{figure}

An important feature of ADfiRe, of particular relevance to the charring regime, is its ability to handle multiple solid fuels with widely different SAV values. This feature is applicable not only to a canopy consisting of a variety of species but also to different parts of the same species, e.g., leaves, twigs, and branches. As the description of the plant characteristics in the experiments of Morandini and Silvani \citep{Morandini_2010} does not contain such details, the potential of ADfiRe in this respect is demonstrated by a simple computation. A fuel mixture is considered, burning under the conditions of $\text{E}_5$, but containing $70\%$ by weight of fines with $\sigma_1=8000\, \text{m}^{-1}$ and $30\%$ coarser combustibles. Figure~\ref{fig:Figure_7}(a) shows the effect on the ROS of the SAV value, $\sigma_2$, of the coarser combustibles and indicates that the thicker the fuel, the slower the ROS. Thus, the slight overprediction of ROS in Figure~\ref{fig:Figure_5a} may be attributed to the neglect of thicker branches in the canopy. Figure~\ref{fig:Figure_7}(b) shows the load of remaining combustibles throughout the field for these cases at $t=$ 200 \text{s} and highlights the differences in consumption as a function of the SAV.  

\section{Conclusions}
    \label{sec:Conclusions}

We present ADfiRe, a wildfire predictive simulation framework based on an advection-diffusion-reaction formalism. The model is enriched with databases from empirical operational tools and with an approximate description of the flow dynamics that benefits from the findings of high-fidelity CFD studies. The core of the model is an energy equation with a control volume extending in height only up to the top of the plantation.

Reaction kinetics include the effects of temperature and oxygen concentration, the latter predicted by estimating air intake to the flame zone and using an appropriate mass balance. Essentially, a single free parameter, $\gamma_r$, is contained in the chemical model, which is tuned to provide fuel consumption for one of the experiments of Morandini and Silvani \citep{Morandini_2010} and gave the correct behavior for all remaining four experiments.

An idea originally proposed by Nelson \citep{Nelson_2002} is extended to model both the role of terrain slope, $\varphi$, and of flame tilting, $\Phi$, due to the wind. The latter angle $\Phi$ is shown to correlate satisfactorily with the Byram number and an empirical relation is used for its prediction. Thus, the overall effect, on fire propagation dynamics, of the flame inclination relative to the ground is incorporated in the model by augmenting the mean gas velocity through the canopy by an appropriately scaled projection of the buoyant velocity vector. This modeling approach resulted in successful prediction of all the fire metrics in the five field experiments, in particular the rate of spread and the fireline intensity. It is noteworthy that these experiments cover a wide range of the spectrum from plume-dominated to wind-dominated fires.

ADfiRe appears to provide a data-efficient and physically interpretable foundation for wildfire modeling that captures the essential macroscopic (and even microscopic) transport mechanisms at a fraction of the computational cost of CFD. However, this approach is based on a heuristic description of the pertinent fluid dynamics by a number of empirical relations, which are motivated by the findings of rigorous CFD studies in the literature. An independent CFD study under the specific conditions of the experimental data of Moradini and Silvani \cite{Morandini_2010}, followed by a detailed computation of the various fluxes and other parameters approximated in ADfiRe,  would be very welcome as a means for further validation and possibly improvement of the present methodology.

Overall, ADfiRe establishes a robust baseline for future data assimilation, uncertainty quantification, and large-scale ensemble forecasting, which are critical capabilities for dynamic fire risk assessment and operational safety management. Ongoing work will interrogate the predictive capabilities of ADfiRe in 2D fire propagation by validating its results against controlled experiments with complex dynamics, particularly merging fire fronts.

\section*{Declaration of Competing Interest}

The authors declare that they have no known competing financial interests or personal relationships that could have appeared to influence the work reported in this paper. 

\section*{Funding Sources}

This research did not receive any specific grant from funding agencies in the public, commercial, or not-for-profit sectors.

\section*{Supplementary data}

The supplementary material submitted with this manuscript contains Appendix A(1-5).

\printcredits

\bibliographystyle{elsarticle-num}
\bibliography{Bibliography}

\end{document}


\let\WriteBookmarks\relax
\def\floatpagepagefraction{1}
\def\textpagefraction{.001}

\shorttitle{ADfiRe: An empirical- and CFD-informed advection-diffusion-reaction thermal model for wildfire spread}

\shortauthors{Vogiatzoglou, et al.,}   

\title[mode = title]{ADfiRe: An empirical- and CFD-informed advection-diffusion-reaction thermal model for wildfire spread}

\section*{Appendix A}
    \label{sec:Appendix}

Equations in Appendix A are numbered with an "S" prefix (e.g., Eq. (S1)) to distinguish them from those in the main manuscript (e.g., Eq. (1)). Same notation for Figures and Tables.

\subsection*{\textbf{A1. Velocity profiles within and above the intact canopy}}
    \label{subsec:Appendix_A1}
    
This analysis is primarily derived from the theoretical framework established by Harman and Finnigan \citep{Finnigan_2007}, with necessary modifications to accommodate the core assumptions of our wildfire simulator.

Assuming a dense and horizontally homogeneous canopy under neutral atmospheric stability \citep{Kiefer_2015, Bebieva_2020}, the canopy penetration depth, $L_c$ [\text{m}], represents the vertical distance required for the atmospheric boundary layer to adjust to the canopy drag as the wind's momentum is absorbed by the foliage. This characteristic length scale is defined as:
\begin{equation}
    \label{eq:Penetration_Depth}
    L_c = \left(\frac{1}{2} \, c_d \, a \, \sigma \right)^{-1},
\end{equation}
where $c_d \approx 0.25$ is the dimensionless leaf-scale drag coefficient, $a$ is the canopy packing ratio, and $\sigma$ [$\text{m}^{-1}$] is the surface-to-volume ratio (SAV) of the fuel bed.

To resolve the complete flow field, the vertical domain is partitioned into the intact canopy region ($z < 0$) and the overlying roughness sublayer ($z > 0$), where $z=0$ is located at the canopy top $H$ [\text{m}]. By employing a mixing-length hypothesis as a turbulence closure model \citep{Finnigan_2000}, the kinematic shear stress is parameterized as $\tau / \rho_g = l_m^{2} \, (du_f/dz)^{2}$, where $\tau$ [$\text{kg} \, \text{m}^{-1} \, \text{s}^{-2}$] is the turbulent shear stress, $\rho_g$ [$\text{kg} \, \text{m}^{-3}$] is the gas density, and $l_m$ [\text{m}] is the height-dependent mixing length. The steady-state, spatially averaged momentum equation within the canopy is then governed by the balance between the vertical divergence of turbulent momentum transport and the canopy drag absorption, and it is given by the following conservation:
\begin{equation}
    \label{eq:Momentum_Balance}
    -\frac{1}{\rho_g} \frac{\partial \tau}{\partial z} - F_D = \frac{d}{dz}\left[l_m^2 \left(\frac{du_f}{dz}\right)^2\right] - \frac{u_f^2}{L_c} = 0,
\end{equation}
where $u_f(z)$ [$\text{m} \, \text{s}^{-1}$] is the spatially and time-averaged horizontal wind speed within the unburned canopy area, and $F_D$ [$\text{m} \, \text{s}^{-2}$] is the kinematic canopy drag. For a sufficiently dense canopy (i.e., a vegetation layer where all momentum is absorbed entirely by the foliage, while exerting zero residual drag on the underlying substrate) the assumptions of a constant mixing length, $l_m$, and constant penetration depth, $L_c$, reduce the momentum balance to an exponential velocity profile, given by:
\begin{equation}
    \label{eq:uf_Inside}
    u_f(z) = u_{f,H} e^{\left(\beta z / l_m\right)}.
\end{equation}
Here, $u_{f,H}$ [$\text{m} \, \text{s}^{-1}$] denotes the mean wind speed exactly at the canopy top ($z=0$), and $\beta = u_* / u_{f,H}$ is the dimensionless velocity ratio, which is approximately $0.3$ under neutral conditions \citep{Finnigan_2007}. The friction velocity, $u_* = \sqrt{\tau(z=0) / \rho_g}$ [$\text{m} \, \text{s}^{-1}$], dictates the momentum transfer from the overlying surface layer. Due to the assumption of constant spatial scaling throughout the dense foliage, the characteristic mixing length relates directly to the penetration depth via:
\begin{equation}
    \label{eq:Mixing_Length}
    l_m = 2 \, \beta^3 \, L_c.
\end{equation}

To capture the flow dynamics above the canopy, we invoke the mixing-layer analogy \citep{Brunet_1994, Raupach_1996}. This hypothesis dictates that the intense shear at the canopy top generates coherent turbulent structures (e.g., Kelvin Helmholtz instabilities), characterized by a vorticity thickness of $l_m / \beta$. Because these canopy-scale eddies dominate the roughness sublayer, extrapolated standard boundary-layer flux-gradient relationships become physically insufficient. Assuming zero vertical divergence of the stress aloft, the standard scaling laws are perturbed via a similarity function, yielding:
\begin{equation}
    \label{eq:Velocity_Gradient}
    \frac{du_f}{dz} = \frac{u_*}{\kappa  (z + d_t)} \Phi_m,
\end{equation}
where $\kappa \approx 0.41$ is the Von Kármán constant and $d_t$ [\text{m}] is the displacement height, equated to the centroid of the canopy drag and is dictated by: 
\begin{equation}
    \label{eq:Displacement_Height}
    d_t = \frac{l_m}{2\beta} = \beta^{2} \, L_c.
\end{equation}

Under neutral atmospheric stability, the modified similarity function simplifies to $\Phi_m = \hat{\phi}_m \left(\beta(z+d_t)/l_m\right)$, effectively embedding the roughness sublayer correction. Integrating Eq.~(S\ref{eq:Velocity_Gradient}) from $z' = -d_t + z_{0m}$ to $z' = z$ provides the above-canopy velocity profile:
\begin{equation}
    \label{eq:uf_Outside}
    \frac{\kappa}{u_*} u_f(z) = \ln\left(\frac{z + d_t}{z_{0m}}\right)
    + \hat{\psi}_m(z),
\end{equation}
where $z_{0m}$ [$\text{m}$] is the roughness length. By enforcing continuity of the wind speed and its gradient at the canopy interface ($z=0$), the roughness length evaluates to $z_{0m} = d_t e^{(-\kappa / \beta)} e^{(\hat{\psi}_m(z=0))}$. The integrated similarity function, $\hat{\psi}_m(z)$, represents the cumulative perturbation of the roughness sublayer and is defined as:
\begin{equation}
    \label{eq:Integrated_Similarity_Function}
    \hat{\psi}_m(z) = \int_{z+ d_t }^{\infty} \left(\frac{1}{z'}\right)
    \left[1 - \hat{\phi}_m \left(\frac{z'}{l_m / \beta} \right)\right]\, dz'.
\end{equation}
Following Harman and Finnigan, the function $\hat{\phi}_m(z)$ quantifies the localized deviation from the standard logarithmic profile and is defined exponentially to ensure asymptotic continuity with the inertial sublayer: 
\begin{equation}
    \label{eq:Similarity_Function}
    \hat{\phi}_m(z) = 1 - c_1 e^{\left(-\beta \, c_2 (z + d_t) / l_m \right)}.
\end{equation}
The coefficient $c_2 \approx 0.5$ relates the vertical scale of the roughness sublayer to the vorticity thickness at the canopy top, while the integration constant is given by $c_1 = \left[1 - \hat{\phi}_m(z=0)\right]e^{\left(c_2/2\right)}$.

To link the ambient meteorological conditions (i.e., $u_w$ measured at 10 \text{m} height) to the internal fuel bed dynamics, Eq.~(S\ref{eq:uf_Outside}) is evaluated at the standard reference height of $z=$ 10 \text{m} alongside the canopy top ($z=$ 0 \text{m}). Solving for the canopy top velocity, $u_{f, H}$, yields:
\begin{equation}
    \label{eq:Canopy_Velocity}
    u_{f, H} = \frac{\kappa u_w}{\kappa + \beta\left[\ln\left((10+d_t)/d_t\right) + \hat{\psi}_m(10) - \hat{\psi}_m(0)\right]},
\end{equation}
where $u_w$ [$\text{m} \, \text{s}^{-1}$] is the ambient reference wind speed. Finally, the bulk shear-induced advection velocity traversing the fuel bed is determined by vertically averaging Eq.~(S\ref{eq:uf_Inside}) over the canopy depth $H$:
\begin{equation}
    \label{eq:Upstream_Velocity}
    u_{f} = \frac{1}{H}\int_{z+ d_t }^{\infty} u_{f,H} \, e^{(\beta z/l_m)} dz = u_{f,H}\left(\frac{l_m}{H \beta}\right).
\end{equation}

\subsection*{\textbf{A2. Two-dimensional extension of the buoyant projection vector}}
    \label{subsec:Appendix_A2}

The wind vector in the horizontal $(x, y)$ plane is defined at an angle $\psi$ relative to the $x$-axis, yielding the horizontal unit vector: $\mathbf{e}_{\mathbf{w}}=\cos\psi\,\mathbf{e}_{\mathbf{x}} + \sin\psi\,\mathbf{e}_{\mathbf{y}}$. We now consider the buoyant velocity vector, $\mathbf{u}_{\textnormal{buoy}}$, which is displaced from the local vertical, $\mathbf{e}_{\mathbf{z}}$, by the deflection angle, $\Phi$ [$^{\circ}$], in the direction of $\mathbf{e}_{\mathbf{w}}$. Decomposing this vector into the vertical component ($z$) and the horizontal wind-aligned component ($w$) yields:
\begin{equation}
    \label{eq:Buoyant_Velocity_2D}
    \mathbf{u}_{\rm{buoy}} = u_{buoy}\left(\sin\Phi \, \mathbf{e}_{\mathbf{w}} + \cos\Phi \, \mathbf{e}_{\mathbf{z}}\right) = u_{buoy} \left(\sin\Phi\cos\psi \, \mathbf{e}_{\mathbf{x}} + \sin\Phi\sin\psi \, \mathbf{e}_{\mathbf{y}} + \cos\Phi \, \mathbf{e}_{\mathbf{z}} \right).
\end{equation}
According to the arguments in Section~2.5 leading to Eq.~(10), the effect of the buoyant velocity on the confined simulation domain is estimated by reducing its magnitude by the scaling factor $u_{log, H}/u_{w}$. Thus:
\begin{equation}
    \label{eq:Downscaling_Buoyant_Velocity}
    \mathbf{u}_{\textnormal{buoy}}' = \left( \frac{u_{\textnormal{log},H}}{u_{w}} \right) \mathbf{u}_{\textnormal{buoy}} \quad \text{and} \quad u_{buoy}' = \left( \frac{u_{\textnormal{log},H}}{u_{w}} \right) u_{buoy},
\end{equation}
where the terms in italics denote the magnitude of the respective vectors.

In the general 2D propagation case, where the wind direction and the gradient of terrain inclination do not coincide, the buoyant velocity contribution, $\mathbf{u}_{\textnormal{proj}}$ [\text{m} \text{s}$^{-1}$], is determined by projecting the vector $\mathbf{u}_{\textnormal{buoy}}'$ onto the plane locally tangent to the terrain. This projection may be readily calculated by subtracting from $\mathbf{u}_{\textnormal{buoy}}'$ the component  that is normal to the tangent plane. Thus:  
\begin{equation}
    \label{eq:Projection_Velocity_2D}
    \mathbf{u}_{\rm{proj}}=\mathbf{u}_{\rm{buoy}}'-\left(\mathbf{u}_{\rm{buoy}}'\cdot\mathbf{n}\right)\mathbf{n},
\end{equation}
where $\mathbf{n}$ is the unit normal to the tangent plane.

For terrain topography defined by the surface function $z = H_t(x, y)$, a differential vector $d\mathbf{r}$ lying on the local tangent plane is given by the expression: 
\begin{equation}
    \label{eq:Diff_Disp} 
    d\mathbf{r} = dx \, \mathbf{e}_{\mathbf{x}} + dy \, \mathbf{e}_{\mathbf{y}} + \left(\frac{\partial H_t}{\partial x}dx + \frac{\partial H_t}{\partial y}dy\right)\mathbf{e}_{\mathbf{z}},
\end{equation}
in terms of the displacements $(dx, dy)$ on the $(x, y)$ horizontal plane. Taking a couple of nonparallel vectors, for example $d\mathbf{r}_x$ and $d\mathbf{r}_y$ with displacements $(dx,0)$ and $(0,dy)$ respectively, and forming their cross product results in the following expression for the unit vector normal to the tangent plane:
\begin{equation}
    \label{eq:Unit_Norm} 
    \mathbf{n} = \frac{1}{\xi} \left(-\frac{\partial H_t}{\partial x} \,  \mathbf{e}_{\mathbf{x}} - \frac{\partial H_t}{\partial y} \, \mathbf{e}_{\mathbf{y}} + \mathbf{e}_{\mathbf{z}}\right) \;,
    \quad \text{where} \quad \xi = \sqrt{\left(\frac{\partial H_t}{\partial x}\right)^2 + \left(\frac{\partial H_t}{\partial y}\right)^2 + 1}.
\end{equation}
Substituting the above formulas in Eq.~(S\ref{eq:Projection_Velocity_2D}) and performing standard algebraic manipulations leads to the following final result for the vector $\mathbf{u}_{\rm{proj}}$:
\begin{equation}
    \label{eq:Projection_2D}
    \begin{split}
    \frac{\mathbf{u}_{\rm{proj}}}{u_{buoy}'} = & \left(\sin\Phi\cos\psi+\frac{\partial H_t}{\partial x}\frac{1}{\xi^2}\frac{\cos(\Phi+\varphi_w)}{\cos\varphi_w}\right) \mathbf{e}_{\mathbf{x}} +
    \left(\sin\Phi\sin\psi+\frac{\partial H_t}{\partial y}\frac{1}{\xi^2}\frac{\cos(\Phi+\varphi_w)}{\cos\varphi_w}\right) \mathbf{e}_{\mathbf{y}} \\
    & +\left(\cos\Phi-\frac{1}{\xi^2}\frac{\cos(\Phi+\varphi_w)}{\cos\varphi_w}\right) \mathbf{e}_{\mathbf{z}} \, ,
    \end{split}
\end{equation}
and for its magnitude:
\begin{equation}
    \label{eq:Magnitude_2D}
    u_{proj}=\left(\frac{u_{\textnormal{log},H}}{u_{w}} \right) u_{buoy} \, \sqrt{1-\frac{1}{\xi^2}\frac{\cos^2(\Phi+\varphi_w)}{\cos^2\varphi_w}}.
\end{equation}
Angle $\varphi_w$ [$^{\circ}$] in Eqs.~(S\ref{eq:Projection_2D}) and (S\ref{eq:Magnitude_2D}) is the inclination of the locally tangent vector $d\mathbf{r}_w$ whose horizontal projection is aligned with the wind direction $\mathbf{e}_{\mathbf{w}} = (\cos\psi, \sin\psi)$. This angle may be determined by the directional derivative of the surface function $H_t(x,y)$ according to the expression:
\begin{equation}
    \label{eq:Tan_Phi_w}
    \tan\varphi_w = \frac{\partial H_t}{\partial x} \, \cos\psi + \frac{\partial H_t}{\partial y} \, \sin\psi.
\end{equation}
It is readily confirmed that, in the case of 1D propagation with constant terrain inclination in the direction of the wind, Eq.~(S\ref{eq:Projection_2D}) gives the simple result
$u_{proj}=\left(u_{log,H}/u_w\right)u_{buoy}(x) \sin(\Phi + \varphi_w)$, where $\varphi_w \equiv \varphi$, i.e. the inclination in the direction of the wind and the maximum terrain inclination, $\varphi$, coincide.

\subsection*{\textbf{A3. Mixed convective heat transfer coefficient}}
    \label{subsec:Appendix_A3}

The mixed convective term in Eq.~(18) combines a natural and a forced convection contribution, $h_{nc}$ and $h_{fc}$, following the classical superposition for combined free–forced convection:
\begin{equation}
    \label{eq:Mixed_Convection}
    U_{conv} = \left( h_{nc}^{3} + h_{fc}^{3} \right)^{1/3}.
\end{equation}

\subsection*{i) Natural convection}
    \label{subsec:Natural_Convection_Appendix_A3}

For a heated horizontal surface losing heat to ambient, the natural convection coefficient is obtained from the following correlation \cite{Bergman_2011}:
\begin{equation}
    \label{eq:Nusselt_nc}
    Nu_{nc} = \frac{h_{nc} L}{k_a} = 0.15\, Gr^{1/3} Pr^{1/3},
\end{equation}
with the Grashof number defined as:
\begin{equation}
    \label{eq:Grashof_Number}
    Gr = \frac{g\,\beta_e\,(T - T_a)\,L^{3}}{\nu^{2}},
\end{equation}
where $k_a$ [$\text{W} \, \text{m}^{-1} \, \text{K}^{-1}$], $\nu$ [$\text{m}^2 \, \text{s}^{-1}$], $\beta_e$ [$\text{K}^{-1}$], and $Pr$ are, respectively, the thermal conductivity, kinematic viscosity, thermal expansion coefficient and Prandtl number of air, evaluated at a reference temperature, and $L$ [$\text{m}$] is a characteristic length (i.e., fireline depth). Substituting $Gr$ into $Nu_{nc}$ and solving for $h_{nc}$, gives:
\begin{equation}
    \label{eq:h_nc}
    h_{nc} = 0.15\, k_a \, Pr^{1/3} \left( \frac{g\,\beta_e}{\nu^{2}} \right)^{1/3} (T - T_a)^{1/3}.
\end{equation}
The characteristic length $L$ cancels identically. Lumping the thermophysical properties into a single coefficient gives the expression used in Eq.~(18):
\begin{equation}
    \label{eq:Natural_Convection}
    h_{nc} = A_{nc}\,(T - T_a)^{1/3}, \qquad 
    A_{nc} = 0.15\, k_a \, Pr^{1/3} \left( \frac{g\,\beta_e}{\nu^{2}} \right)^{1/3},
\end{equation}
with $A_{nc} \, [\mathrm{W\,m^{-2}\,K^{-4/3}}]$ evaluated at the reference air temperature and reported in Table 1.

\subsection*{ii) Forced convection}
    \label{subsec:Forced_Convection_Appendix_A3}

The forced convection coefficient is driven by the intra-canopy shear velocity $u_{shear}$ [$\text{m} \, \text{s}^{-1}$] at the top of the fuel layer ($z=H$), through the correlation \cite{Churchill_1983}:
\begin{equation}
    \label{eq:Nusselt_fc}
    Nu_{fc} = \frac{h_{fc} L}{k_a} = 0.17\, Re^{0.8} Pr^{0.33},
\end{equation}
with the Reynolds number based on the same characteristic length $L$ used for the mass-transfer analogy in Eq.~(3):
\begin{equation}
    \label{eq:Reynolds_Number}
    Re = \frac{u_{shear}(H)\, L}{\nu}.
\end{equation}
Substituting and solving for $h_{fc}$, gives:
\begin{equation}
    \label{eq:h_fc}
    h_{fc} = 0.17\, k_a \, Pr^{0.33} \, \nu^{-0.8} \, L^{-0.2} \, u_{shear}(H)^{0.8}.
\end{equation}
Lumping the thermophysical properties as before gives the working expression used in Eq.~(18):
\begin{equation}
    \label{eq:Forced_Convection}
    h_{fc} = A_{fc}\, u_{shear}(H)^{0.8}\, L^{-0.2}, \qquad
A_{fc} = 0.17\, k_a\, Pr^{0.33}\, \nu^{-0.8},
\end{equation}
with $A_{fc} \, [\mathrm{W\,m^{-2.6}\,K^{-1}\,s^{0.8}}]$ evaluated at the reference air temperature and reported also in Table 1. 

\subsection*{\textbf{A4. Additional macroscopic quantities of interest}}
    \label{subsec:Appendix_A4}

\begin{figure}
    \centering
    \includegraphics[width=.9\textwidth]{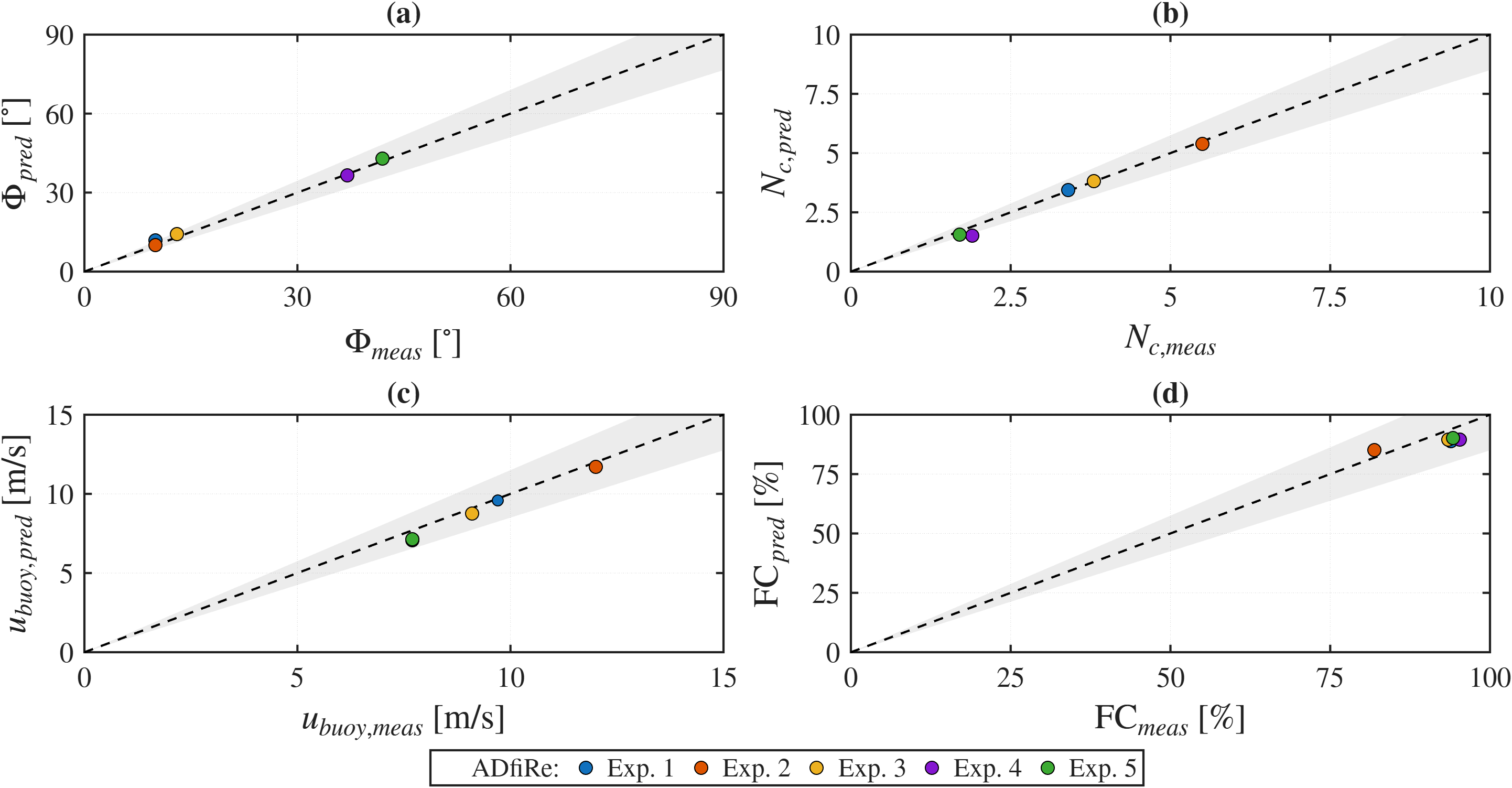}
    \caption{Validation of ADfiRe against empirical fire metrics from \cite{Nelson_2015, Morandini_2010}. Parity plots compare predicted vs. measured: (a) flame tilt angle - $\Phi$, (b) Byram number - $N_c$, (c) buoyant velocity - $u_{buoy}$, (d) fuel consumption ratio - FC. Dashed line indicates 1:1 agreement, and shaded area is $\pm 15\%$ tolerance. Colors denote campaigns (Exp. 1–5).}
    \label{fig:Figure_5b}
\end{figure}

The metrics depicted in Figure~S\ref{fig:Figure_5b} are resolved directly by ADfiRe's coupled physics-based formulation and cannot be evaluated via conventional semi-empirical models (such as Rothermel). The predicted quantities of interest present a good agreement compared to the observational data of Morandini \& Silvani \cite{Morandini_2010}. The agreement in the Byram number ($N_c$) resulting from the simulations to the values calculated from the experimental data by Nelson \cite{Nelson_2015} (Figure~S\ref{fig:Figure_5b}(b)) is particularly encouraging, as it indicates that the model satisfactorily describes both plume- and wind-dominated fire fronts \cite{Morvan_2018, Trouve_2020, Morvan_2025, Moinuddin_2021}.

\subsection*{\textbf{A5. Mesh sensitivity analysis}}
    \label{subsec:Appendix_A5}

\begin{figure}
    \centering
    \includegraphics[width=.6\textwidth]{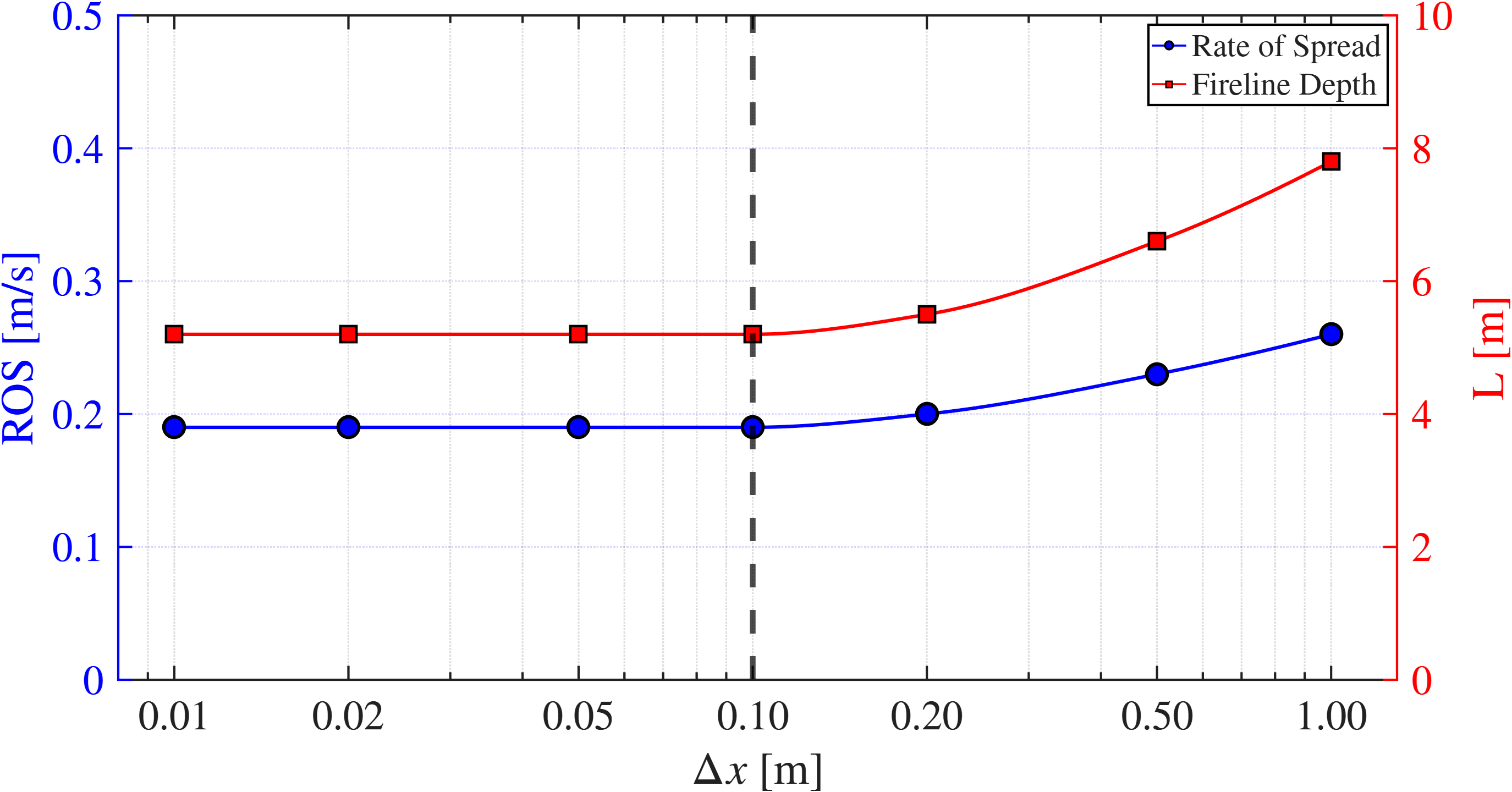}
    \caption{Grid sensitivity analysis for Experiment 1: Rate of spread (ROS, left axis, blue line with circles) and fireline depth ($L$, right axis, red line with squares) as a function of the spatial grid resolution, $\Delta_x$ [m]. Markers denote discrete simulation results across the seven grid levels ($\Delta_x =$ 0.01 m - 1.00 m), connected by smooth continuous interpolations. The vertical dashed line highlights the baseline discretization ($\Delta_x =$ 0.10 m) adopted in the study, illustrating asymptotic convergence for $\Delta_x \leq$ 0.10 m and the onset of numerical diffusion for coarser meshes ($\Delta_x \geq$ 0.20 m).}
    \label{fig:Figure_8}
\end{figure}

To assess numerical convergence, a grid sensitivity study was conducted across seven spatial discretizations ($\Delta_x =$ 0.01 m, 0.02 m, 0.05 m, 0.10 m - base, 0.20 m, 0.50 m, and 1.00 m), evaluating the rate of spread (ROS) and fireline depth ($L$) against the fine reference resolution ($\Delta_x =$ 0.01 m). As illustrated in Figure~S\ref{fig:Figure_8} , asymptotic convergence is achieved for all resolutions $\Delta_x \leq$ 0.10 m with relative error less than 1.00\% in both ROS and $L$, whereas coarser grids $\Delta_x \geq$ 0.20 m introduce numerical diffusion that overpredicts ROS and $L$. Although the finer resolutions ($\Delta_x =$ 0.01 m, 0.02 m) demand substantially longer execution times due to the large node counts, they yield almost identical predictions to $\Delta_x =$ 0.10 m, used in our simulations. Consequently, the baseline mesh of $\Delta_x =$ 0.10 m provides the optimal trade-off, ensuring fully converged physical fidelity at minimal computational cost.

Table~S\ref{table:Mesh_Sensitivity} presents the mesh sensitivity analysis and compares the two quantities of interest against the reference solution, indicating also the corresponding computational time.

\begin{table}
    \centering
    \caption{Spatial mesh sensitivity and grid convergence analysis for Experiment 1 (E1). Each simulation was executed, using a single CPU core on a laptop (AMD Ryzen 7 5800H, 3.20 GHz, 16 GB RAM).}
    \label{table:Mesh_Sensitivity}
    \renewcommand{\arraystretch}{1.5}
    \begin{tabular}{cccccccc}
    \toprule
    \textbf{Grid Level} & $\Delta x$ \textbf{[m]} & \textbf{Nodes} $N_x$ & \textbf{ROS [m/s]} & $\varepsilon_{\mathrm{ROS}}$ \textbf{[\%]} & $L$ \textbf{[m]} & $\varepsilon_L$ \textbf{[\%]} & \textbf{Execution Time} \\
    \midrule
    Grid 1 (Ref.) & 0.01 & 20001 & 0.190 & Reference & 5.20 & Reference & $\sim 387 \, \text{min}$ \\
    Grid 2 & 0.02 & 10001 & 0.191 & 0.53 & 5.20 & 0.10 & $\sim 83 \, \text{min}$ \\
    Grid 3 & 0.05 & 4001 & 0.191 & 0.55 & 5.21 & 0.19 & $\sim 6 \, \text{min}$ \\
    Grid 4 (Base) & 0.10 & 2001 & 0.192 & 1.05 & 5.21 & 0.21 & $\sim 1 \, \text{min}$ \\
    Grid 5 & 0.20 &  1001 & 0.201 & 5.79 & 5.51 & 5.96 & $\sim 0.85 \, \text{min}$ \\
    Grid 6 & 0.50 & 401 & 0.235 & 23.68 & 6.64 & 27.62 & $\sim 0.49 \, \text{min}$ \\
    Grid 7 & 1.00 & 201 & 0.264 & 38.94 & 7.83 & 50.57 & $\sim 0.41 \, \text{min}$ \\
    \bottomrule
    \end{tabular}
\end{table}

\vspace{1cm}

\bibliographystyle{elsarticle-num}
\bibliography{Bibliography}